\documentclass[aps,prl,twocolumn,floatfix,superscriptaddress]{revtex4}

\usepackage{latexsym} 
\usepackage{graphicx}
\usepackage{rotating}
\usepackage{hyperref}
\usepackage{amsmath,amssymb,amsfonts}
\usepackage{bm}
\usepackage{color}

\newcommand{\smallwidth}{0.86\columnwidth}
\newcommand{\figwidth}{1.0\columnwidth}

\newcommand{\miniwidth}{0.7\columnwidth}

\newcommand{\dx}{d$_{x2-y2}$ }
\newcommand{\psig}{p$_\sigma$ }
\newcommand{\dz}{d$_{3z2-r2}$ }
\newcommand{\pz}{p$_{\pm z}$ }
\newcommand{\px}{p$_{(x,y)}$ }

\newcommand{\lasco}{La$_{2}$CuO$_4$ }

\newcommand{\ncco}{Nd$_2$CuO$_4$ }

\newcommand{\emphasize}{\emph}

\def\onlinecite#1{\cite{#1}}

\newcommand{\up}{\uparrow}
\newcommand{\dn}{\downarrow}

\newcommand{\vk}{\bold{k}}

\begin{document}

\title{Apical oxygens and correlation strength in electron and hole doped copper oxides}
\author{C\'edric Weber}
\author{Kristjan Haule}
\author{Gabriel Kotliar}
\affiliation{Department of Physics, Rutgers University,  Piscataway, NJ 08854, USA}

%%%%%%%%%%%%%%%%%%%%%%%%%%%%%%%%%%%%%%%%%%%%%%%%%%%%%%%%%%%%%%%%
%%%%%%%%%%%%%%%%%%%%%%%%%%%%%%%%%%%%%%%%%%%%%%%%%%%%%%%%%%%%%%%%
%%%%%%%%%%%%%%%%%%%%%%%%%%%%%%%%%%%%%%%%%%%%%%%%%%%%%%%%%%%%%%%%
%%%%%%%%%%%%%%%%%%%%%%%%%%%%%%%%%%%%%%%%%%%%%%%%%%%%%%%%%%%%%%%%
%%%%%%%%%%%%%%%%%%%%%%%%%%%%%%%%%%%%%%%%%%%%%%%%%%%%%%%%%%%%%%%%

 \begin{abstract}
   We use the Local Density Approximation in combination with the Dynamical Mean Field Theory to carry out a comparative
   investigation of a typical electron doped and a typical hole doped copper
   oxide, NCCO and LSCO respectively. The parent compounds of both  materials are strongly correlated electron
   systems in the vicinity of the metal to charge transfer insulator transition.
   In NCCO the magnetic long range order is essential to  open a charge transfer gap, while Mott physics
   is responsible for the gap in LSCO. We highlights the role of the apical oxygens in determining the strength of
   the correlations and obtaining overall good agreement between theory and several experimentally determined
   quantities. Results for optical conductivity, polarized X-ray absorption and angle resolved photoemission
   are presented and compared with experiments.
 \end{abstract}

 \maketitle

%%%%%%%%%%%%%%%%%%%%%%%%%%%%%%%%%%%%%%%%%%%%%%%%%%%%%%%%%%%%%%%%
%%%%%%%%%%%%%%%%%%%%%%%%%%%%%%%%%%%%%%%%%%%%%%%%%%%%%%%%%%%%%%%%
%%%%%%%%%%%%%%%%%%%%%%%%%%%%%%%%%%%%%%%%%%%%%%%%%%%%%%%%%%%%%%%%
%%%%%%%%%%%%%%%%%%%%%%%%%%%%%%%%%%%%%%%%%%%%%%%%%%%%%%%%%%%%%%%%
%%%%%%%%%%%%%%%%%%%%%%%%%%%%%%%%%%%%%%%%%%%%%%%%%%%%%%%%%%%%%%%%
%%%%%%%%%%%%%%%%%%%%%%%%%%%%%%%%%%%%%%%%%%%%%%%%%%%%%%%%%%%%%%%%
%%%%%%%%%%%%%%%%%%%%%%%%%%%%%%%%%%%%%%%%%%%%%%%%%%%%%%%%%%%%%%%%
%%%%%%%%%%%%%%%%%%%%%%%%%%%%%%%%%%%%%%%%%%%%%%%%%%%%%%%%%%%%%%%%
%%%%%%%%%%%%%%%%%%%%%%%%%%%%%%%%%%%%%%%%%%%%%%%%%%%%%%%%%%%%%%%%
%%%%%%%%%%%%%%%%%%%%%%%%%%%%%%%%%%%%%%%%%%%%%%%%%%%%%%%%%%%%%%%%

\section{\bf Introduction}

Since their discovery, the electronic structure of the high
temperature superconductors has been a subject of intensive
theoretical attention as well as controversy, a situation that
continues even today. A landmark question to understand 
these materials is how their physical properties follow from their electronic
structure and to which extent simplified descriptions in the form
of model Hamiltonians describe their basic physical properties
in the normal state.

It is generally accepted that the physics of the copper oxide
based high temperature superconductor families is captured by the
copper-oxygen layers and the relevant degrees of freedom are the copper
$d_{x^2-y^2}$ orbitals and the oxygen p$_x$ and p$_y$
orbitals \cite{emery_3bands_model,varma_3bands_model}. Numerous
studies have demonstrated that this model captures 
qualitatively the physics of the copper oxide planes. However, 
there have been several proposals that the \pz oxygen orbitals and the \dz
copper orbital play an important role in the onset of orbital current
order \cite{orbital_cedric} or for the existence of
superconductivity \cite{apical_BCS,weber_BCS}. 

Seeking a simpler low energy description, several studies 
\onlinecite{zhang_rice_singlet,zhang_singlet}
have shown that the Hubbard model describes some {\it
qualitative} properties of the copper oxygen layers. However,
the precise energy range over which the description is valid and
the quality of this description for different physical
observables, is still a subject of active research.

A related question is hence how to map the copper oxide layers onto the
various effective Hamiltonians and what is the effective strength
of the Coulomb correlations in these systems. 
There are various approaches tackling this issue, 
ranging from ab-initio methods to model
Hamiltonian studies. 

The theoretical studies usually consist of a two stage
process: In the first step
ab-initio approaches use an approximate
technique, such as constrained DFT in the LDA or GGA
approximation, or quantum chemical methods, to derive the
parameters and the form of the effective Hamiltonian
\onlinecite{macmahan_parameters,hybertsen,more_ref7}. 
In the second step, model Hamiltonian based approaches
compute various observables in the framework of a
given model Hamiltonian and given approximation technique.
The parameters entering the model Hamiltonians are determined
by comparing the results side by side with experiment.

Numerous research efforts notwithstanding, even the basic question
of the strength of the correlations in the copper oxide materials
and the origin of the insulating gap in their parent compound
has not been fully elucidated. There are two opposing physical
pictures describing the origin of the insulating gap in these materials.
In the so called Slater picture, the insulating behavior is understood as
the result of a doubling of the unit cell caused by
antiferromagnetic long range order. 
In the so called Mott picture, the
insulating behavior is the result of the local blocking of the
electron propagation due to the strong Coulomb repulsion.
In the latter picture, the insulating behavior originates from the localization of the
electron and is not tied to any specific form of magnetic long range order. 
Hence antiferromangetic long range order arises as a secondary
instability. In the presence of magnetic long range order, the
unit cell is doubled and the two pictures, Slater and Mott, are
continuously connected: No physical observable can provide a
sharp distinction between the two, both magnetic order and
blocking contribute to the insulating behavior.
Mean field theory treatments allow to study the paramagnetic state 
as an underlying {\it normal state} mean field solution, 
which can supports a sharp transition between a paramagnetic metal and a 
paramagnetic insulator. 
This solution is not realized when other more stable mean field
solutions supporting long range order exist, but within a mean field
framework, it can still be used to draw a sharp distinction between
Slater insulators and Mott insulators, by
investigating if the ordered state is derived from a metallic or
an insulting paramagnetic solution.
The pioneering work of Zaanen Sawatzky Allen \cite{zaanen} and
their sharp (so called ZSA) boundary between charge transfer metals and
charge transfer insulators can be viewed in this light.

A simple argument can be formulated for the Hubbard model to estimate
the strength of the correlation of cuprates.
Within a one band Hubbard description, parametrized by a bandwidth $W \approx 8 t$ 
and a Coulomb repulsion U, the insulating gap of the
paramagnetic insulator is $U-W$ and the super-exchange is $J=4 t^2/ U$.
For cuprates, the gap is around $\approx 2$eV and $J \approx 0.1$eV,
and therefore it is found that the repulsion is of the order $U/W \approx 1.5$, 
which is above but not far from the Mott boundary $U \approx W$, and hence 
the cuprates are in a regime of intermediate correlation strength.
Conclusions on the placement of cuprates
in a regime of intermediate correlation strength were also
reached by numerical studies (for a review see for example Ref.~\onlinecite{more_ref12}).

The strength of the correlations was also studied in the three band theory.
In particular, large N slave boson mean field theory of a three band
model (with no oxygen-oxygen transfer integrals) \cite{sarma_mean_field} 
of the copper oxides obtained a sharp transition between
the metal and the charge transfer insulator in the paramagnetic phase.
This metal-to-charge-transfer-insulator transition
parallels the Brinkman Rice transition in the Hubbard model.
It was found that there is indeed a correspondence between the two critical Coulomb
repulsions of the Hubbard model $U_{c1}$ (the minimal Coulomb repulsion
that supports the paramagnetic insulator) and $U_{c2}$ (the minimal
repulsion that does not support a paramagnetic metal) with the two
critical charge transfer energies $\Delta_{c1}$ and $\Delta_{c2}$, where the paramagnetic
insulating state and the paramagnetic metallic state are destroyed,
respectively. The critical value of the charge
transfer energy in the three band theory ($\Delta_{c2}$) plays the role of the critical U ($U_{c2}$)
of the Hubbard model \cite{sarma_mean_field}.
Although a strong particle-hole asymmetry is expected in the three band theories,
since doped electrons reside mainly on copper sites while doped
holes reside mainly on the oxygen sites, it was shown that the resulting
quasiparticle band structure was surprisingly particle-hole
symmetric. Indeed, it is due to the strong copper-oxygen hybridization
which results in the formation of Zhang-Rice singlets and to the quasiparticles
that involves copper spin and oxygen charge. 

Other more realistic treatments of the three band description of the copper oxides were carried out within the slave boson
framework. Some include the oxygen dispersion
\cite{grilli_kotliar_millis} \cite{auerbach}, additional copper
and oxygen orbitals \onlinecite{feiner}, short range magnetic correlations \cite{more_ref11,schmalian_three_band_slave_boson_1}, 
the nearest neighbor Coulomb interactions \onlinecite{raimondi} or the electron phonon coupling \onlinecite{grilli_castellani_three_band}. 
Within slave boson mean field theory of the three band model, 
the parent compound of hole doped cuprates LSCO
was located  close  but above (i.e. on the insulating side of) 
the metal to charge transfer insulator transition boundary \cite{review_gabi}.

The development of Dynamical Mean Field Theory 
\cite{OLD_GABIS_REVIEW} and their extensions 
\cite{more_ref13,our_review,vcpt_senechal} opened new avenues to advance
our qualitative understanding of the electronic structure of the cuprates
and its quantitative description. 
DMFT goes beyond slave bosons theories: This method treats both coherent and incoherent features on the
same footing, whereas slave bosons theories are not able to capture the coherent character 
of the physical solution. DMFT successfully describes the Mott transition
of the Hubbard model and gives a 
deeper understanding of the Brinkman Rice transition \cite{OLD_GABIS_REVIEW,uc2_def}.
Single site DMFT calculations can also be extended to more precise cluster calculations.
In particular, cluster corrections in DMFT allow to assess the validity of the single site calculations. 

The question of the strength of the correlations was also addressed by DMFT
studies of multi-band model Hamiltonians for the copper 
oxides planes \cite{dmft_paper_krauth_gabi,avignon,dmft_paper_krauth_gabi}.
The phase diagram, with respect to the charge transfer energy and the Coulomb repulsion
of the copper orbitals was studied.
The boundary between the metallic and the charge-transfer-insulator solutions 
was located, as well as the crossover line between the charge-transfer-insulator and the Mott insulator.
A full phase diagram of a copper oxide model, and a very complete
analogy with the DMFT studies of the Mott transition in the
Hubbard model was also performed recently \cite{luca_long_paper}. 

Combination of DMFT with electronic structure methods, such as LDA,
allow to combine the ab-initio and the model Hamiltonian viewpoint in the LDA+DMFT framework.
The LDA+DMFT method \cite{our_review} allow in particular 
to determine the strength of correlations 
for specific materials, like NCCO and LSCO, and there are still several 
important issues unresolved regarding how this picture is connected to the cuprates:
a) How should the different materials be placed in the
qualitative ZSA phase diagram. Should the parent compounds of the
copper oxide materials be thought as Slater or Mott/charge
transfer Insulators, 
b) What significant differences are there in the different level of description, mainly
what are the differences between the one band and three band theory,
c) What is the quality of the description of the various experimental 
observables,
for the different low energy models involving a different number of bands,
and finally 
d) Can one obtain a consistent picture of the spectroscopies of hole and electron
doped cuprates using a first principles method.

In regards to point a), the issue is still under debate. 
Using the analysis of model Hamiltonians,
Refs \cite{luca_nature,luca_long_paper} classify the parent
compounds of electron and hole doped compounds 
as Slater insulators in the metallic side of the ZSA phase diagrams.
Previous LDA+DMFT studies 
classified LSCO \cite{our_previous_paper_lsco} as a Mott insulator (or more precisely as a charge transfer insulator),
and NCCO \onlinecite{our_nature_paper} was identified as Slater insulator. 
On the other hand, the first principles study of \cite{more_ref6} found  NCCO's parent compound to be Mott insulator. 
Finally, the phenomenological analysis of experimental data in Ref \cite{more_ref1} 
concludes that the analysis of the optical data
requires comparable strength of interactions in hole and electron doped cuprates.

In regards to point b) the energy range over which the spectral functions of the three band model are reproduced by the one
band model is also subject of controversy. 
For example, Refs. \cite{luca_nature,luca_long_paper} assert
that the three band model and a one band model of the copper oxides are quantitatively equivalent up to a scale of 4eV, 
while a similar analysis by A. Macridin et al. concludes that the range of validity of the Hubbard model is much smaller,
and is of the order of 0.5 eV \cite{macridin_1band_validity}. 
An additional controversy, regarding the differences between the one
band and the three band theory, is connected to the values of the
insulting gap in the paramagnetic insulator. In particular, in a one
band theory near $U_{c2}$, the gap of the paramagnetic solution is
substantially smaller then the gap of the antiferromagnetic solution
for the same parameters \cite{rosenberg_1band,millis_compound_at_uc2,luca_nature}. 
This is not the case in the
three band model where antiferromagnetism increases the value of the
paramagnetic insulating gap by less than $15\%$ \cite{our_previous_paper_lsco}. The latter
statement is controversial with respect to Ref \cite{luca_long_paper}, which argued that
the one band model and the three band model are similar in their
physics up to energies as large as twice the gap.

Regarding point c), many authors used the theoretical results obtained with the Hubbard model
to fit experiments. For instance Ref. \cite{kyung_U_dependant} considered a treatment of the
Hubbard model with variational cluster perturbation theory (VCPT).
It could match experiments but it required  
a sensible dependence of the Hubbard U on the level of doping.
On the other hand, an approximate
diagrammatic treatment of the one band model \cite{bansil1} indicates that the
experimental optical spectra and the dispersion \cite{more_ref3,more_ref4} 
can be reproduced without a doping dependent U.

Note that a three band LDA+DMFT study was able to describe the experimental data
of NCCO without having to invoke doping dependent parameters
\onlinecite{our_nature_paper}. 
Moreover, using the same technique, a successful 
description of both the integrated optical weights and
the magnitude of the optical conductivity below the
charge transfer gap for LSCO and NCCO was obtained 
\cite{our_previous_paper_lsco}.
This approach however does not give the correct magnitude of the optical conductivity of 
LSCO for energies of the order of the charge transfer gap, 
suggesting that additional orbitals might play an important role in LSCO.

In this paper we reconsider these issues building on our earlier
work of Refs \onlinecite{our_previous_paper_lsco} and
\onlinecite{our_nature_paper}. We use an ab-initio approach, e.g.
the Local Density Approximation combined with the Dynamical Mean
Field Theory (LDA+DMFT) \cite{our_review} to study the electronic
structure of NCCO and LSCO. The comparative study of two
typical cuprate compounds LSCO and NCCO allow us to place firm
bounds ascertaining the importance of correlations in the
cuprates. The good agreement between theory and experiments,
achieved using single site LDA+DMFT within a multiband framework,
for two different compounds, is a significant results and
illustrates the power of this new first principles approach to correlated materials. 
Our results contrast with a recent phenomenological analysis of optical data, of electron and doped cuprates:
They concluded that for a one band theory, vertex corrections
beyond single site DMFT were required to obtain a reasonable fit to experiments \cite{more_ref1}.

In this work, we include also the apical oxygens and an additional
copper orbital (\dz), which were not included in our previous study of
LSCO \cite{our_previous_paper_lsco}. This extends the
quantitative agreement between theory and experiments to a
broader energy range. It sheds light on why the three band model
description of NCCO \onlinecite{our_nature_paper} agrees with experiments up to larger energy scales than for
LSCO \onlinecite{our_previous_paper_lsco}. NCCO, in the T' structure, lacks
apical oxygens and is therefore well described by the three band
model up to a much higher energy scale, justifying the excellent
agreement between theory and experiment found in Ref \cite{our_nature_paper}.

We confirm that for integration cutoffs smaller than half the
gap of the parent compound, the additional
apical oxygen  degrees of freedom do not affect the integrated
optical weights in LSCO. This validates the analysis carried out
in Ref \cite{our_previous_paper_lsco} to extract the strength of
correlations in these materials, which was based on analysis up to
an energy scale of half the charge transfer gap, i.e. 1 eV. The
apical oxygens however substantially affects the shape of the
optical conductivity and it strongly modifies the optical
conductivity of LSCO around 2 eV. While the hole occupancy of
the $p_z$ orbital is small, it has a clear effect on the optical
conductivity of the hole doped cuprates.
We show here that within the first principles LDA+DMFT framework,
extensions of the model to include further orbitals or longer range
correlations (using cluster extension) consistently improves precision
of the calculation and improves agreement with experiments.

The organization of our paper is the following.
In section 1 we describe the phase diagram of NCCO and LSCO within single site and two site
cluster DMFT, and highlight the role of magnetic order and
singlet correlations in these materials. In section 2 we present
the evolution of the angle resolved photoemission spectra (ARPES) of
these materials, stressing the various features of the theory
that require a description beyond static mean field theory and
its comparison with the  experiments. 
Details regarding the formula used in section 2 to compare the spectral weight
with experiments are given in Appendix A. 
Section 3 focuses on the optics and how the evolution of the optical properties with
doping and temperature in NCCO and LSCO can be understood as the
result of placing the two compound on two different sides of the ZSA boundary, once
the band structure of both materials is taken into account. 
The inclusion of the apical oxygens in the theoretical modeling is
important to obtain correct results for various physical quantities,
including the shape of the optical conductivity
and the integrated optical spectral weights with a
cutoff of the order of the charge transfer gap. We also discuss
the connection of the optical conductivity with various features
in the ARPES spectra, and compute the occupation of the different
orbitals which are relevant to the XAS spectra. There is one
parameter, the double counting correction, whose determination in
the LDA+DMFT approach is not unique. We thus examined in the last
section the dependence of our results on this parameter and used
this dependence to estimate the proximity of both NCCO and LSCO
to the ZSA boundary. The tight binding parametrization obtained
makes contact with model Hamiltonian studies and is reported in
Appendix B. We conclude with some outlook for further work.

%%%%%%%%%%%%%%%%%%%%%%%%%%%%%%%%%%%%%%%%%%%%%%%%%%%%%%%%%%%%%%%%
%%%%%%%%%%%%%%%%%%%%%%%%%%%%%%%%%%%%%%%%%%%%%%%%%%%%%%%%%%%%%%%%
%%%%%%%%%%%%%%%%%%%%%%%%%%%%%%%%%%%%%%%%%%%%%%%%%%%%%%%%%%%%%%%%
%%%%%%%%%%%%%%%%%%%%%%%%%%%%%%%%%%%%%%%%%%%%%%%%%%%%%%%%%%%%%%%%
%%%%%%%%%%%%%%%%%%%%%%%%%%%%%%%%%%%%%%%%%%%%%%%%%%%%%%%%%%%%%%%%
%%%%%%%%%%%%%%%%%%%%%%%%%%%%%%%%%%%%%%%%%%%%%%%%%%%%%%%%%%%%%%%%
%%%%%%%%%%%%%%%%%%%%%%%%%%%%%%%%%%%%%%%%%%%%%%%%%%%%%%%%%%%%%%%%
%%%%%%%%%%%%%%%%%%%%%%%%%%%%%%%%%%%%%%%%%%%%%%%%%%%%%%%%%%%%%%%%
%%%%%%%%%%%%%%%%%%%%%%%%%%%%%%%%%%%%%%%%%%%%%%%%%%%%%%%%%%%%%%%%
%%%%%%%%%%%%%%%%%%%%%%%%%%%%%%%%%%%%%%%%%%%%%%%%%%%%%%%%%%%%%%%%

\section{\bf Formalism}

LDA+DMFT uses first principles density functional theory methods
to extract the hopping parameters of the model, which is subsequently
solved using DMFT and its extensions. The LDA calculation was carried out with  the PWSCF package
\cite{cappuccion}, which employs a plane-wave basis set and
ultrasoft pseudopotentials \cite{lda_vanderbilt}. Downfolding to
a three band model, containing copper \dx and two oxygen \psig
orbitals was performed by the maximally localized Wannier
functions (MLWF) method \cite{lda_basis,lda_basis2}. The
downfolded LDA band structure of \ncco (NCCO) and \lasco (LSCO) (see Appendix B) results in the
following three band Hamiltonian:
\begin{multline}
  \label{eq:3band_hub1}
  \mathcal{H}_t = \sum\limits_{ij\sigma, (\alpha,\beta) \in (p_x,p_y,d_{x2-y2})}{  t^{\alpha \beta}_{ij} c^\dagger_{i \alpha \sigma} c_{j \beta \sigma} } \\
                + \epsilon_p \sum_{i \sigma \alpha \in (p_x,p_y)}{\hat n_{i \alpha \sigma}} +
                  \left(\epsilon_d-E^{dc}\right) \sum_{i\sigma}{\hat n_{i d \sigma}}
\end{multline}
where $i$ and $j$ label the CuO$_2$ unit cells of the lattice, and
$t_{ij}^{\alpha\beta}$ are the hopping matrix elements.
$\epsilon_d$ and $\epsilon_p$ are the on-site energies of the $d$ and
$p$ orbitals, respectively.
Finally, we note that the charge transfer energy $\epsilon_d-\epsilon_p$
plays the same role as U in the single band Hubbard model,
as seen for example in slave bosons mean-field studies \cite{review_gabi}.

To this Hamiltonian, we add the onsite Coulomb repulsion $U$ on the
\dx orbital
\begin{equation}
\label{eq:3band_hub2}
\mathcal{H}_U = U_d \sum_{i}{ \hat n_{id\up} \hat n_{id\dn}}
\end{equation}
where the value of $U_d=8 eV$. The LDA+DMFT method, accounts for
the correlations which are included in both LDA and DMFT by a
double counting correction to the $d$-orbital energy, $E_{dc}=U_d(n_d-0.5)$,
which amounts to a shift of the relative positions of the d and p orbitals.
Here we take $n_d$ to be the occupancy of the correlated orbital in the parent compound,
which gives the double counting corrections $E_{dc}=4.8eV$ ($3.12eV$) for NCCO (LSCO).

The LDA downfolded parameters are shown in table \ref{table:lda_param},
which we find to be close to those extracted by other first
principles methods. 
An extended 6-band model, that considers the $d_{3z2-r2}$, $p_z$ and $p_{-z}$ orbitals, was also considered
for LSCO. The 6-band Hamiltonian is:  
\begin{multline}
 \label{eq:6band_hub1}
  \mathcal{H}_{apical}=\sum\limits_{ij\sigma, (\alpha,\beta) \in (p_x,p_y,p_{\pm z},d_{x2-y2},d_{3z2-r2})}
               {  t^{\alpha \beta}_{ij} c^\dagger_{i \alpha \sigma} c_{j \beta \sigma} } + \\
                  \sum_{i\sigma, \alpha \in (p_x,p_y,p_{\pm z})}{\epsilon_{\alpha} \hat n_{i \alpha \sigma}} +
                  \sum_{i\sigma, \alpha \in (d_{x2-y2},d_{3z2-r2})}  {   \left(\epsilon_{\alpha}-E^{dc}\right)   \hat n_{i\alpha\sigma}} + \\
           U_{d}  \left( \sum_{i,       \alpha \in (d_{x2-y2},d_{3z2-r2})}  {  \hat n_{i\alpha\up} \hat n_{i\alpha\dn} }  + 
                         \sum_{i}                                           {\hat n_{i d_{x2-y2}} \hat n_{i d_{3z2-r2}} } \right) 
\end{multline}
The hopping parameters $t_{ij}$ were obtained by downfolding the LDA band structure to 
six orbitals ($d_{x2-y2}$,$d_{3z2-r2}$,$p_{x,y}$,$p_{z,-z}$).
The double counting for the 6-band model 
is defined as $E_{dc} = U_d \left( n_{d_{x2-y2}} + n_{d_{3z2-r2}} - 0.5 \right) $. 
The same on-site repulsion $U_d$ was considered for the \dx and \dz orbitals.

\begin{table}
\caption{ LDA band calculations gives us
different set of parameters for LSCO ,NCCO and PCCO compounds.
The f states of the Nd and Pr atoms have been treated
as core states and are not treated as valence state. Long-range
hopping (not shown) are also considered within the calculations.
The amplitude of the nearest neighbors hoppings
($t_{pp}$,$t_{dp}$), the LDA on-site energies
($\epsilon_p^0$,$\epsilon_d^0$) and the on-site repulsion $U_d$ are shown in this table in eV. 
In electron notations the bonding orbitals enter the Hamiltonian
with a negative transfer integral sign, and the anti-bonding
orbitals with a positive sign.}
\begin{tabular}{|c|c|c|c|c|}
  \hline
  Compound & $\epsilon_d^0-\epsilon_p^0$ [eV] & $U_d$ [eV] & $t_{dp}$ [eV] & $t_{pp}$ [eV] \\
  \hline
    NCCO (This work) &  1.61 & 8  & 1.16 & 0.54
  \\
    PCCO (This work) &  1.65 & 8  & 1.17 & 0.54
  \\
    LSCO (This work) &  2.76 & 8 & 1.41 & 0.66
  \\
  \hline
  \hline
    NCCO (Ref.~\onlinecite{gavrichkov}) &  1.42 & 10  & 1.18  & 0.69
  \\
    LSCO (Ref.~\onlinecite{gavrichkov}) & 0.918  &  10  & 1.357  & 0.841
  \\
    LSCO (Ref.~\onlinecite{macmahan_parameters})  & 3.5 & 7.9   & 1.5  & 0.6 
  \\
    LSCO (Ref.~\onlinecite{hybertsen}) & 3.6 & 10.5  & 1.3   & 0.65 
  \\
  \hline
  \hline
\end{tabular}
\label{table:lda_param}
\end{table}

We solve these models using Dynamical Mean Field Theory (DMFT), in which the Green's function is given by:
\footnote{All calculations have been carried out at temperature $T=89^\circ K$ when not specified.}.
 \begin{equation}
   \label{greenfunc}
   \textbf{G}_\vk(i \omega_n) =  \left( i \omega_n + \mu - \bold{H}_\vk - \bold{\Sigma}(i \omega_n)   \right)^{-1},
 \end{equation}
where $\bold{H}_\vk$ is the Fourier transform of the $\mathcal{H}_t$ in
Eq.~(\ref{eq:3band_hub1}) and Eq.~(\ref{eq:6band_hub1}).
$\bold{\Sigma}$ is the self-energy matrix being nonzero only in the $d$ orbital.
The self energy in Eq.~(\ref{greenfunc}) is obtained
by solving an Anderson impurity model subject to the  DMFT
self-consistency condition:
\begin{equation}
 \left(  i\omega - E_{imp} - \bold\Sigma(i\omega) - \bold\Delta(i\omega) \right)  = \hat P \left( \frac{1}{N_k} \sum\limits_{\vk \in BZ}  {G_\vk (i \omega)} \right) ^{-1},
\end{equation}
where the sum runs over the first Brillouin Zone (BZ), and $\hat P$ is projecting the averaged
green function onto the impurity cluster subspace. 

In this work we use the continuous time quantum Monte Carlo impurity solver algorithm
\cite{Haule_long_paper_CTQMC,werner_ctqmc_algorithm}.
Real frequency resolved quantities were obtained by analytic
continuation of the observables obtained in matsubara frequencies. We have crossed checked the 
analytic continuation of the observables obtained on the matsubara axis
using several other numerical solvers: the exact diagonalization solver (ED)~\cite{OLD_GABIS_REVIEW}, 
the density matrix renormalization group solver (DMRG) \cite{amaricci_dmrg}, 
and the NCA solver (Non-crossing diagram approximation).
CTQMC and ED/DMRG/NCA are complementary tools, working respectively on the matsubara
and real axis. 

Magnetism was considered within the single site DMFT by 
solving two independent impurity problems, 
while in the case of the 2-site cluster DMFT (c-DMFT), the 2-site magnetic unitcell 
is mapped to a 2-impurity cluster. Cluster DMFT improves single site DMFT by
adding the non-local self energy, not present in the single site DMFT.
The 6-band calculation maps the 2-correlated orbitals \dx and \dz to 
an impurity containing two different orbitals.

Finally, model Hamiltonian (\ref{eq:3band_hub1}) and (\ref{eq:3band_hub2}) were
studied previously (for examples see Refs.~\onlinecite{dmft_paper_krauth_gabi} 
and \onlinecite{Avignon_DMFT_perovskite}).
When $U_d$ is large, there is a metal to charge transfer insulator
transition at integer filling, as a function of the charge
transfer energy $\epsilon_d-\epsilon_p$. However the electron
doped cuprates fall in a regime where the phases with magnetic
long range order occupy a large fraction of phase space and this
regime was not investigated  previously.

%%%%%%%%%%%%%%%%%%%%%%%%%%%%%%%%%%%%%%%%%%%%%%%%%%%%%%%%%%%%%%%%
%%%%%%%%%%%%%%%%%%%%%%%%%%%%%%%%%%%%%%%%%%%%%%%%%%%%%%%%%%%%%%%%
%%%%%%%%%%%%%%%%%%%%%%%%%%%%%%%%%%%%%%%%%%%%%%%%%%%%%%%%%%%%%%%%
%%%%%%%%%%%%%%%%%%%%%%%%%%%%%%%%%%%%%%%%%%%%%%%%%%%%%%%%%%%%%%%%
%%%%%%%%%%%%%%%%%%%%%%%%%%%%%%%%%%%%%%%%%%%%%%%%%%%%%%%%%%%%%%%%
%%%%%%%%%%%%%%%%%%%%%%%%%%%%%%%%%%%%%%%%%%%%%%%%%%%%%%%%%%%%%%%%
%%%%%%%%%%%%%%%%%%%%%%%%%%%%%%%%%%%%%%%%%%%%%%%%%%%%%%%%%%%%%%%%
%%%%%%%%%%%%%%%%%%%%%%%%%%%%%%%%%%%%%%%%%%%%%%%%%%%%%%%%%%%%%%%%
%%%%%%%%%%%%%%%%%%%%%%%%%%%%%%%%%%%%%%%%%%%%%%%%%%%%%%%%%%%%%%%%
%%%%%%%%%%%%%%%%%%%%%%%%%%%%%%%%%%%%%%%%%%%%%%%%%%%%%%%%%%%%%%%%
%%%%%%%%%%%%%%%%%%%%%%%%%%%%%%%%%%%%%%%%%%%%%%%%%%%%%%%%%%%%%%%%

\section{Phase diagram of LSCO and NCCO}

In this section we discuss the phase diagram of NCCO and LSCO, and in particular their magnetic
properties. We compare the various treatments of the short-range and long-range correlations. 
In particular, we consider the ordered state within the single site DMFT and within the 
2-site cluster DMFT.

\begin{figure}
\begin{center}
\includegraphics[width=\figwidth]{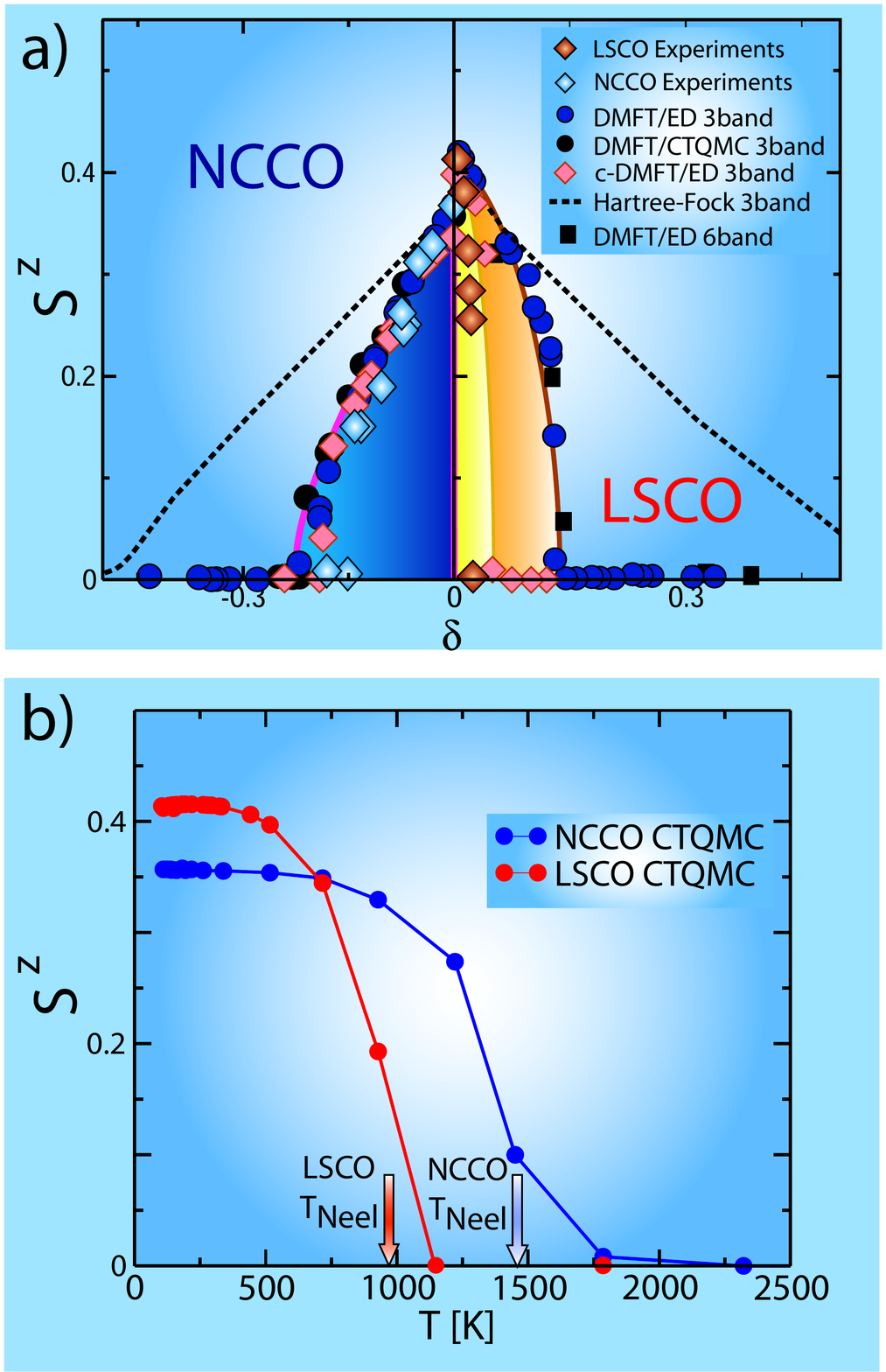}
\caption{
(Colors online)
a) We show the theoretical mean-value of the staggered
magnetization $S^z=\frac{1}{2}\left( n_{\uparrow} - n_{\downarrow} \right) $
obtained for LSCO and NCCO by
both single site DMFT (LDA+DMFT) and cluster cellular DMFT (LDA+cDMFT) of the 3-band theory, and DMFT of the 6-band theory.
Experimental values $M(\delta)/M_0$ ($M_0=M(\delta=0)$) for NCCO \cite{staggered_moment} and LSCO \cite{lsco_magnetic_moment} are also shown,
and for comparison with DMFT, we assume $M_0=M_{DMFT}(\delta=0)$, where $M_{DMFT}$ is the magnetic moment at 0 doping obtained
by single site DMFT.
The general trend compare well to our data, and for NCCO we obtain a qualitative agreement for the critical doping.
For LSCO we find that the magnetization vanish at large doping within the single site DMFT approximation, but when
the possibility of local singlet is allowed (cDMFT) the range of stabilization of the long-range order is much reduced.
Note that the theoretical magnetic moment was obtained at constant temperature $T=89^\circ K$, and experimentally at
$2^\circ K$ for LSCO and $8^\circ K$ for NCCO.
b) Magnetization versus temperature in the parent compounds of NCCO and LSCO
obtained by single site DMFT for the three band theory. The extracted mean-field Neel temperature is about 1500K for NCCO and 1000K for LSCO.
}
\label{fig:magnetism}
\end{center}
\end{figure}

In Fig.~\ref{fig:magnetism}{\bf a} we show the magnetic moments of NCCO obtained 
within a three-band description (left panel of Fig.~\ref{fig:magnetism}).
Single site DMFT data of NCCO (blue circles, left panel) are in remarkable agreement with experimental data for all dopings 
\cite{staggered_moment}, 
though LDA+DMFT slightly overestimates the range of stabilization of the Neel state:
in experiment, the magnetic order of NCCO vanishes at  
doping $\delta \approx 0.15$, while in our calculations we find a slightly 
larger critical doping of $\delta \approx 0.2$.
We also show data for the cluster cellular DMFT (LDA+cDMFT) (grey diamonds). 
For NCCO, the c-DMFT and the single site DMFT
data are almost identical. This is a very strong test that 
the magnetic correlations of NCCO are well captured by single site DMFT.
In the case of LSCO (right panel of Fig.~\ref{fig:magnetism}), we found significant difference in the region of stability 
of the magnetic state between the single site and cluster DMFT ($\delta<10\%$), 
hence the dynamical short range correlations - absent in single site approach - 
are very important for LSCO in the underdoped regime, as reported in \cite{schmalian_three_band_slave_boson_1}.
For LSCO, we also considered the 6-band theory (the 6-band calculations include the description 
of the \pz and \dz orbitals), which does not change magnetic moment of the three band 
theory. Experimental values for LSCO \cite{lsco_magnetic_moment} are also shown 
and compare qualitatively to our data.

We also carried out the Hartree Fock calculation, and we found that in
this static approach the magnetization vanishes only at unrealistic
 large doping $\delta \approx 50\%$ for both NCCO and LSCO 
(dashed line in Fig.~\ref{fig:magnetism}{\bf a}), 
which points towards the important role of dynamic
correlations at finite electron and hole doping.
Indeed, in the Hartree-Fock calculations, the picture is fairly simple: 
The magnetic state is stabilized due to an optimization of the
Coulomb energy at the expense of the kinetic energy (the staggered
magnetic order avoids double occupation) and in this picture there is
no strong difference between LSCO and NCCO, as shown in Fig.~\ref{fig:magnetism}{\bf a})
The same mechanism is responsible for slight overestimation of the
critical doping within DMFT.

When comparing to experiments, it is important to keep in mind
that two dimensional compounds are not able to sustain infinite-range
magnetic order at a finite temperature (Mermin-Wagner theorem
\cite{mermin_wagner}). Therefore, the Neel temperature within DMFT
should be interpreted as the temperature below which the magnetic
correlations become long but remain finite. This temperature can be
much higher then the actual Neel temperature of the material, which is
controlled by the magnetic exchange between the two dimensional copper
oxide layers; and vanishes for a well separated copper oxide planes.
In Fig.~\ref{fig:magnetism}{\bf b} we show the temperature dependence of the magnetic
moment for the 3-band calculations of LSCO and NCCO.
The extracted mean-field Neel temperature is about 1500K for NCCO and 1000K for LSCO,
which is much higher than the experiment Neel temperature.
Finally, we emphasize that the magnetic moment of the parent compound of LSCO $m=0.42$
is larger than the one of NCCO $m=0.35$, which might suggests that LSCO is more correlated
than NCCO.
\begin{figure}
\begin{center}
\includegraphics[width=\figwidth]{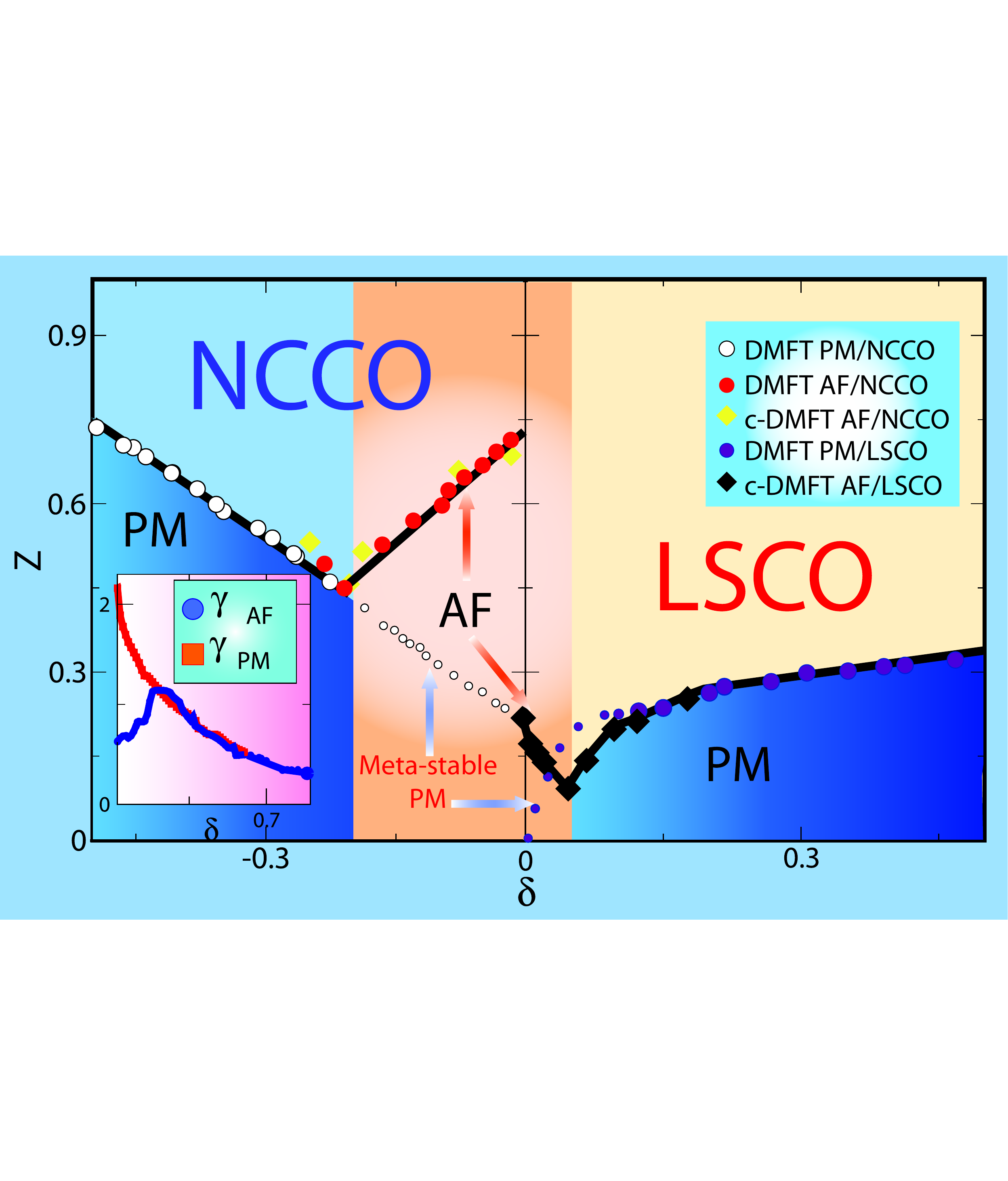}
\caption{
(Colors online) We show the quasi-particle weight Z for the 3-band description of LSCO and NCCO.
The quasi particle weight of the paramagnet (PM) is finite at integer filling for NCCO, which is a signature that the
paramagnetic state of NCCO is a metal.
In the ordered state of NCCO (AF), the quasi-particle weight
is estimated by the specific heat of the A and B magnetic sublattices $\gamma_{i} \propto \frac{\rho_i(\epsilon_{F})}{Z_i}$ and
$i=A,B$, and the total quasi-particle weight Z is given by
$Z =  \sum\limits_i{\rho_i(\epsilon_F)}  / \sum\limits_i{ \frac{\rho_i(\epsilon_{F})}{Z_i} }$.
The specific heat $\gamma$ of the AF and PM states of NCCO are shown in the inset.
}
\label{fig:z}
\end{center}
\end{figure}
The strength of the correlations can also be quantized 
by calculating the quasi-particle renormalization amplitude $Z$ (see Fig.~\ref{fig:z}).
First, we consider the paramagnetic solutions of both NCCO (white circles on the left side) 
and LSCO (blue circles on the right side).
We find that the quasi-particle weight is going to zero in the parent compound of LSCO: This is
a signature that LSCO is a charge transfer insulator. For NCCO, we find that it is going to a finite value at zero
doping, which is the signature that NCCO is a paramagnetic metal.

At finite doping, we find that the quasi-particle weight of 
the paramagnetic state is much larger in NCCO than in LSCO. This is a signature that
NCCO is more metallic and less correlated than LSCO.

We now turn to the calculations for the ordered state of both LSCO (red circles on the left) and NCCO (red diamonds on the right).
Here the average quasi-particle weight in the ordered state 
was obtained by the following formula:
$Z = \sum\limits_i{\rho_i(\epsilon_F)}  / \sum\limits_i{ \frac{\rho_i(\epsilon_{F})}{Z_i} }$.
The motivations comes from the formula for the specific heat of 
the magnetic system given by $\gamma_{i} \propto \frac{\rho_i}{Z_i}$ with $i=A,B$.

The quasi-particle weight in the ordered state of NCCO is larger than the one for LSCO,
which shows that the ordered state of NCCO is also less correlated than LSCO, 
and hence the character of the paramagnetic
state of the parent compounds (paramagnetic metal versus paramagnetic insulator) 
has direct consequences for the magnetic solutions.

We find that the mechanism to open a gap at integer filling is totally different for hole and electron doped
compounds, and our results place NCCO and LSCO to different regions of the ZSA phase diagram.
For hole doped compound NCCO, the quasi-particles are scattered increasingly and get a
very short lifetime when approaching the insulator (charge transfer insulator).
In the electron doped compound NCCO, the system 
minimizes its free energy by doubling the unit-cell which opens a Slater gap (Slater insulator).

Finally, we note that the presence of magnetism is concomitant with an entropy loss
, which results in an increase of the quasi-particle weight.
The direct consequence is that the specific heat $\gamma$ of
the ordered state is lower than the specific heat of the paramagnet (inset of Fig.~\ref{fig:z}).

%%%%%%%%%%%%%%%%%%%%%%%%%%%%%%%%%%%%%%%%%%%%%%%%%%%%%%%%%%%%%%%%
%%%%%%%%%%%%%%%%%%%%%%%%%%%%%%%%%%%%%%%%%%%%%%%%%%%%%%%%%%%%%%%%
%%%%%%%%%%%%%%%%%%%%%%%%%%%%%%%%%%%%%%%%%%%%%%%%%%%%%%%%%%%%%%%%
%%%%%%%%%%%%%%%%%%%%%%%%%%%%%%%%%%%%%%%%%%%%%%%%%%%%%%%%%%%%%%%%
%%%%%%%%%%%%%%%%%%%%%%%%%%%%%%%%%%%%%%%%%%%%%%%%%%%%%%%%%%%%%%%%
%%%%%%%%%%%%%%%%%%%%%%%%%%%%%%%%%%%%%%%%%%%%%%%%%%%%%%%%%%%%%%%%
%%%%%%%%%%%%%%%%%%%%%%%%%%%%%%%%%%%%%%%%%%%%%%%%%%%%%%%%%%%%%%%%
%%%%%%%%%%%%%%%%%%%%%%%%%%%%%%%%%%%%%%%%%%%%%%%%%%%%%%%%%%%%%%%%
%%%%%%%%%%%%%%%%%%%%%%%%%%%%%%%%%%%%%%%%%%%%%%%%%%%%%%%%%%%%%%%%
%%%%%%%%%%%%%%%%%%%%%%%%%%%%%%%%%%%%%%%%%%%%%%%%%%%%%%%%%%%%%%%%
%%%%%%%%%%%%%%%%%%%%%%%%%%%%%%%%%%%%%%%%%%%%%%%%%%%%%%%%%%%%%%%%

\clearpage

\section{Photoemission and Fermi Surface of NCCO}

The basic quantity describing the electronic structure of the
material is the electronic spectral function:
\begin{equation}
A^\alpha(\bold{k},\omega)=-\frac{ \left( \textrm{Im} \bold{G}(\bold{k},\omega)^{\alpha,\alpha} \right)}{\pi}
\end{equation}
Where $\alpha$ is the orbital index, and $\bold{k}$ is running through the unfolded Brillouin zone. 
The total spectral weight is $A(\bold{k},\omega)=\sum\limits_{\alpha}{A^\alpha(\bold{k},\omega)}$.
Experiments like Angle-Resolved Photo-Emission Spectroscopy (ARPES) are able to probe the k-dependent spectral functions,
and can therefore be compared side-by-side with theoretical calculations \footnote{
For better comparison with experiments, we mapped the ordered state calculations
$A^\alpha(\bold{K},\omega)$ to the paramagnetic Brillouin zone, see Appendix A.}.
In this section we investigate the agreement between the theoretical spectra
and ARPES measurements.

\begin{figure}
\begin{center}
\includegraphics[width=\figwidth]{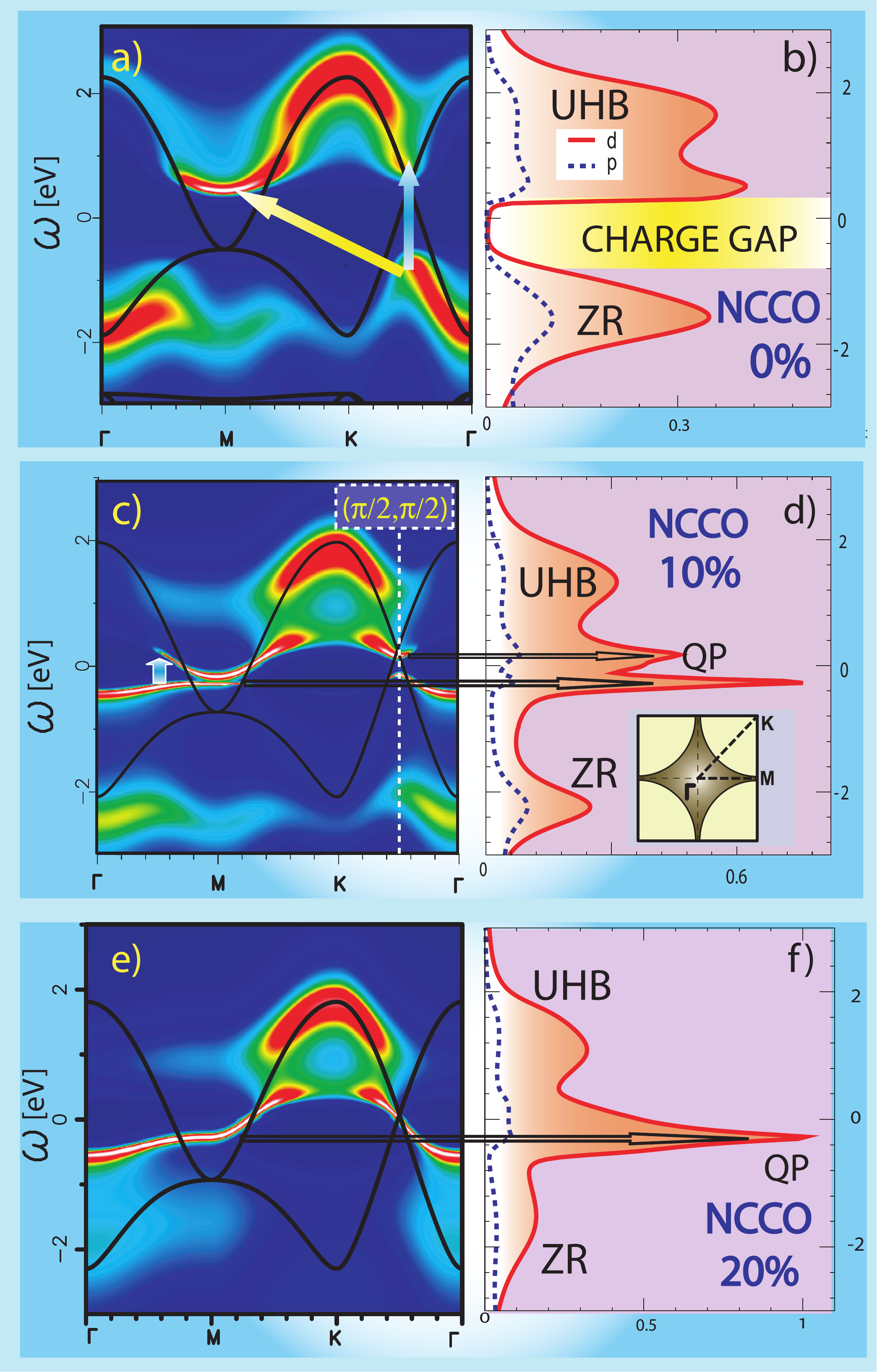}
\caption{
(Colors online) 
Frequency dependent spectral weight A(k,w) obtained by LDA+DMFT of NCCO at
a) integer-filling, c) $10\%$ and e) $20\%$ electron doping for NCCO. A(k,w) was obtained
along the usual path $\Gamma-M-K-\Gamma$ in the Brillouin zone (see inset of d). The solid lines are the rigid LDA bands.
The partial density of states of the d and p orbitals are shown in b),d),f).
The density of states is showing the upper Hubbard band (UHB), the Zhang rice singlet (ZR) and the quasi-particle peak (QP) for the doped compounds.
a) At integer-filling, we find for NCCO that the indirect gap is $\approx 1.2$eV, between $(\pi,0)$ and $(\pi/2,\pi/2)$,
as shown by the diagonal arrow, and the direct gap is about $1.5$eV, as indicated by the vertical arrow.
d) At $10\%$ doping we observe a splitting of the quasi-particle peak due to magnetism. The splitting
of the quasi-particle peak is associated with the magnetic pseudo-gap at $(\pi/2,\pi/2)$ in panel (c).
The horizontal arrows pointing from c) to d) are guide to the eyes.
The optical transitions at $10\%$ doping occur around $(\pi,0)$ within the quasi-particle band, as indicated by the vertical arrow.
e) At $20\%$ doping, magnetism is destroyed and the pseudo-gap at $(\pi/2,\pi/2)$ is closed. The quasi-particle
peak in panel (f) is clearly related to the spectral weight close to the Fermi surface at $(\pi,0)$, as indicated by
the horizontal arrow. 
}
\label{fig:DOS}
\end{center}
\end{figure}

In Fig.~\ref{fig:DOS}{\bf a,c,e} we show the spectral functions resolved in momentum space. In Fig.~\ref{fig:DOS}{\bf b,d,f} we show the integrated
spectral functions, that show the energy locations of the main spectral features.

In the parent compound (Fig.~\ref{fig:DOS}{\bf a-b}), 
we find two dispersive peaks separated by the charge transfer gap
of about $1.2eV$, as expected in NCCO \cite{ncco_charge_gap}. 

The spectral feature below the Fermi level (Fig.~\ref{fig:DOS}{\bf b}) 
is an admixture of oxygens and copper orbitals
, commonly known as the Zhang-Rice singlet (ZR). It is worth noting that our results show that
the oxygens orbitals carry no magnetic moment. The oxygen sites
average the magnetization on its both copper neighbors.

When the Slater gap opens, there is a spectral weight transfer from the upper
Hubbard band to the Zhang-Rice singlet and the 
lower Hubbard band (located at $-10$eV, not shown), such that 
minority spectral weight is concentrated in the upper Hubbard band (UHB), and
the majority spectral weight is mostly present in the lower Hubbard band (LHB) and
in the Zhang-Rice singlet (ZR).

The top of the lower band occurs at $(\pi/2,\pi/2)$, while the bottom
of upper band appears at $M=(\pi,0)$, therefore the gap is indirect (see
yellow arrow in panel \textbf{a}).
Those two bands can also be
obtained in the simpler Hartree Fock approximation, though the size of
the gap is overestimated in a static mean-field.

At $10\%$ electron doping (Fig.~\ref{fig:DOS}{\bf c-d}), NCCO is still magnetic, and therefore
the Zhang-Rice singlet and the upper Hubbard band are well separated. Those
two features are also observed by the simpler Hartree Fock approximation. 
What is clearly not visible in static mean-field, is the 
presence of a very sharp and narrow band slightly below and above the Fermi level (Fig.~\ref{fig:DOS}{\bf c}),
that corresponds to the quasi-particle peak (QP) in the integrated spectra (Fig.~\ref{fig:DOS}{\bf d}).
It is worth noting that the optical transitions occur at this doping within these narrow bands,
from the narrow band below the Fermi level to the narrow band above the Fermi level, as depicted
in Fig.~\ref{fig:DOS}{\bf c} by the vertical arrow.
In the ordered state of NCCO we observe in the ordered phase a splitting of
the quasi-particle peak (Fig.~\ref{fig:DOS}{\bf d}) into two structures. The first 
corresponds to the narrow band bellow the Fermi level, and its
main weight is at $M=(\pi,0)$, as indicated by the lower horizontal arrow.
The peak slightly above the Fermi
level is due to the spectral weight at $\bold{k}=(\pi/2,\pi/2)$ (see upper horizontal arrow),
and is related to the pseudo-gap around $\bold{k}=(\pi/2,\pi/2)$.

Upon larger doping $20\%$ (Fig.~\ref{fig:DOS}{\bf e-f}), magnetism disappears and 
the pseudo-gap at $\bold k=(\pi/2,\pi/2)$ closes. The peak slightly above
the Fermi level, that was present at $10\%$ doping, now disappears. 

\begin{figure}
\begin{center}
\includegraphics[width=\figwidth]{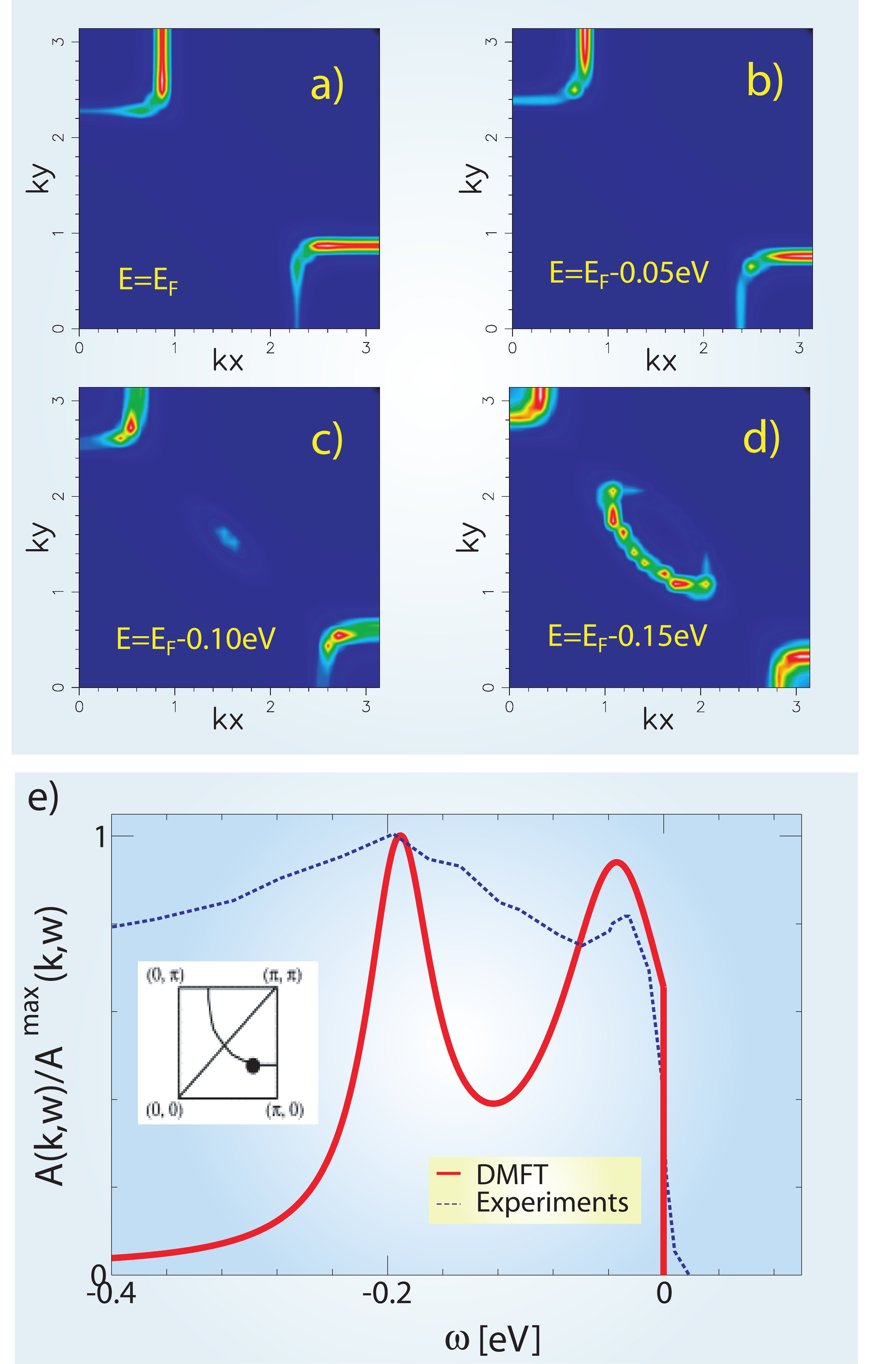}
\caption{
(Colors online) 
Fermi surface maps obtained by DMFT calculations in the ordered state at $10\%$ doping,
a) at the Fermi energy and b)-d) at lower energies ranging from -0.05eV to -0.15eV.
The Fermi surface map at the Fermi level is shaped by the presence of magnetism, whereas the arc in
the energy map at energy -0.15eV is also present in the paramagnetic calculations. 
e) comparison of $A(\omega)$ for a fixed k point $k=(3\pi/4,\pi/4)$
(shown in the inset) obtained theoretically (lower curve) and
experimentally from Ref.~\cite{arpes_comparison} (upper curve). The peak at -0.03eV
is due to magnetism, and the peak at lower energy -0.2eV is associated to the arcs 
seen in d).
}
\label{fig:fse}
\end{center}
\end{figure}

\begin{figure}
\begin{center}
\includegraphics[width=\figwidth]{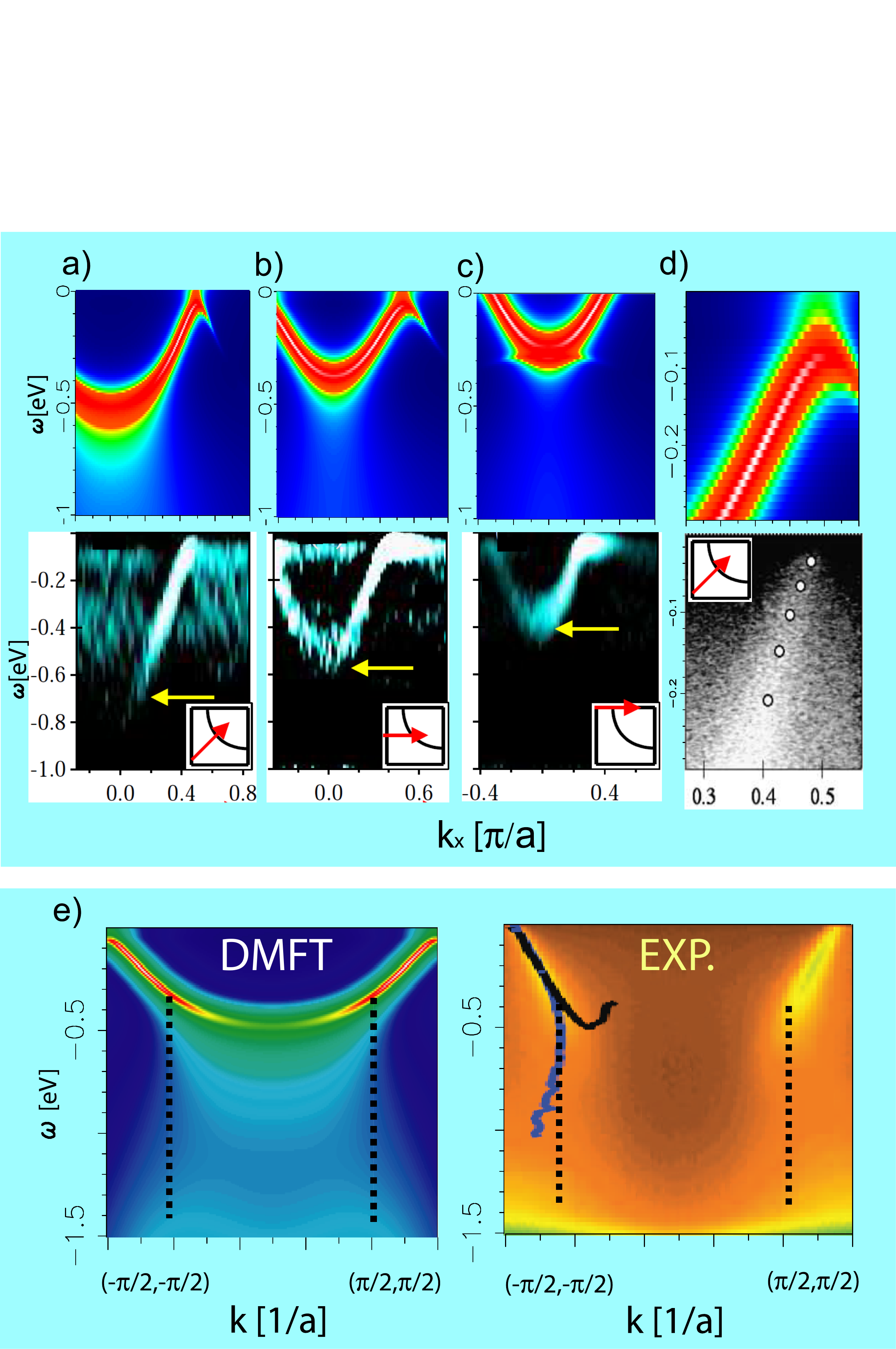}
\caption{
(Colors online) 
Side by side comparison of $A(k,\omega)$, along the path as depicted in the inset, obtained
theoretically (upper row) and experimentally (lower row), 
a)-c) at $15\%$ electron doping 
from Ref.~\onlinecite{ikeda_kinks}, and 
d) at $13\%$ doping from Ref.~\cite{arpes_comparison}.
e) Comparison between DMFT (left side) and experiments of Ref.~\onlinecite{moritz_waterfall_ncco} (right side), along
the nodal cut of the Brillouin Zone at $17\%$ electron doping.
The agreement between DMFT and experiments is remarkable. The vertical dashed lines are guide to the eyes to
illustrate the presence of a sharp kink in the dispersion (\emphasize{waterfall}). 
}
\label{fig:path}
\end{center}
\end{figure}

\begin{figure}
\begin{center}
\includegraphics[width=\figwidth]{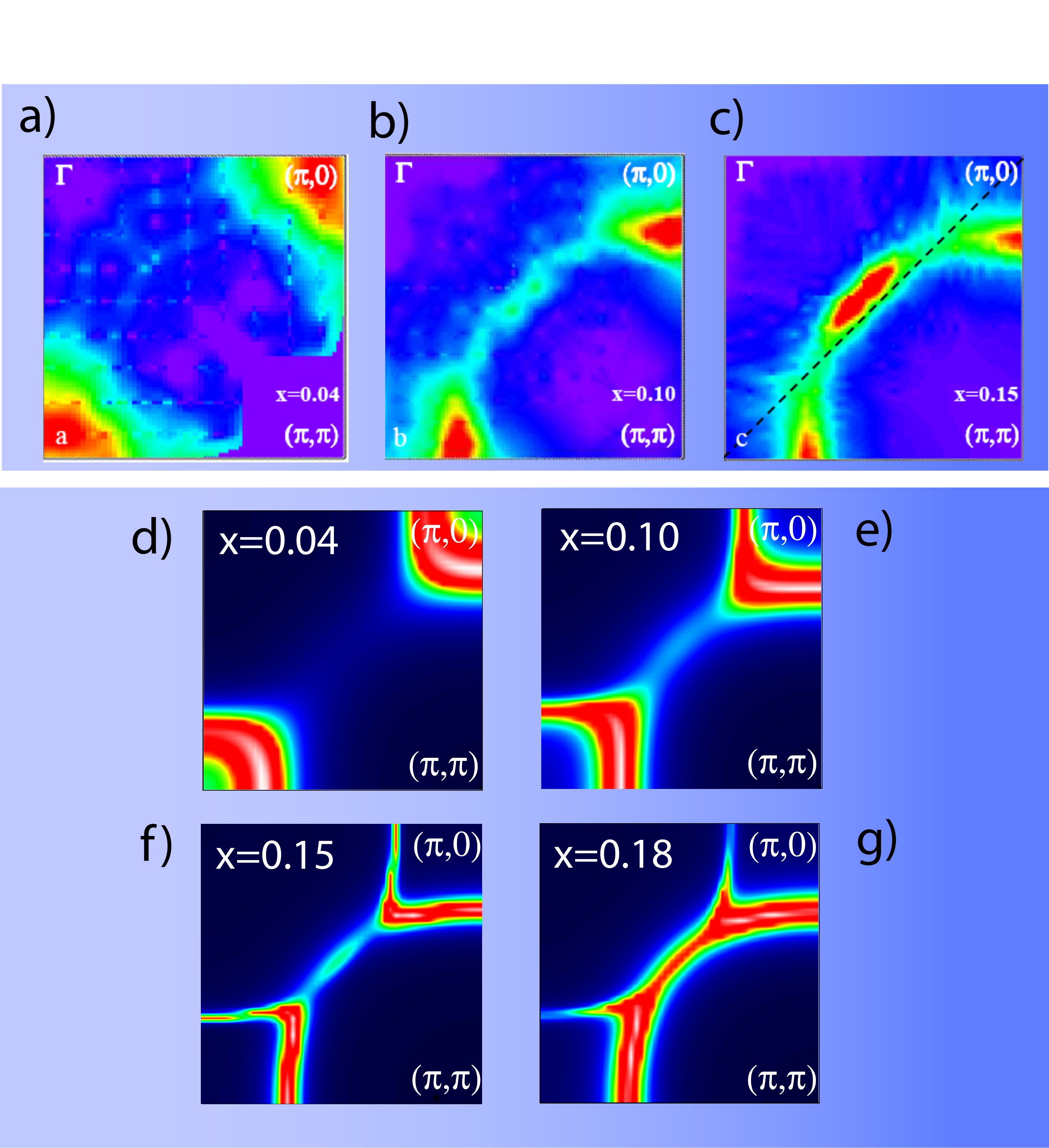}
\caption{
(Colors online) Side by side comparison of the
Fermi surface obtained experimentally (reproduced from
Ref.~\onlinecite{armitage_fermi_prl}) and obtained by single site DMFT
calculations for NCCO in the ordered state. a) Experimental
results for doping $4\%$, b) for $10\%$ electron doping
and c) $15\%$ electron doping. The results are compared to DMFT calculations
for similar dopings d)-g). The Fermi surface at low
doping (a) is centered around $M=(\pi,0)$, and is moving towards the
Fermi arc shape (g) when magnetism is destroyed.
}
\label{fig:fsdop}
\end{center}
\end{figure}

We now turn to Fig.~\ref{fig:fse}{\bf a-d}, where we show 
the spectral functions resolved in momentum
space at fixed energies $E_f$, $E_f-0.05$eV, $E_f-0.1$eV and $E_f-0.15$eV, at 
finite doping $10\%$. At this doping NCCO is magnetic. 
The magnetic Fermi surface in panel (\textbf{a}) has
a square-like shape structure centered around $M=(\pi,0)$. At lower
energy, in panel \textbf{d}, we observe the presence of an arc centered around
$\bold{k}=(\pi/2,\pi/2)$.
This comes primarily from the pseudo-gap around momentum point $k=(\pi/2,\pi/2)$,
which is a signature of the magnetic long range order (the
Fermi surface of the ordered state is gapped at $k=(\pi/2,\pi/2)$.
The Fermi surface of the ordered state moves towards the usual Fermi arc 
shape when the system becomes metallic and the pseudogap at $k=(\pi/2,\pi/2)$ is closed
in the paramagnet.

Some aspects of the doped electronic structure can be understood in
terms of the Hartree Fock rigid band picture, for example the holes
appear first upon doping at the $M=(\pi,0)$ point, 
but the renormalization of the bands, and the multiple peak structure
in energy for a given momentum point (see Fig.~\ref{fig:fse}\textbf{e}),
are not captured in static mean-field. 

In Fig.~\ref{fig:fse}\textbf{e} we show the energy dependence of the
spectral function at a fixed k point (shown in the inset of the figure). 
The peak close to the Fermi
energy is connected to the square-like 
Fermi surface of panel (\textbf{a}) and is hence connected to magnetism.
The peak at lower energy $-0.2$eV,
is related to the arc shape of panel \textbf{d} and has paramagnetic character.
The peak positions are in a very good
agreement with recent angle resolved photo-emission measurements of
Ref.~\onlinecite{arpes_comparison} also shown in Fig.~\ref{fig:fse}\textbf{e}.

In Fig.~\ref{fig:path}{\bf a-c} we compare side by side experimental
data (middle panels) \cite{ikeda_kinks} and DMFT calculations
(upper panels). The agreement is quantitative showing that
our approach captures the low energy physics of NCCO.
The pseudo-gap at $(\pi/2,\pi/2)$ is also observed in experiments \cite{arpes_comparison}
(Fig.~\ref{fig:path}\textbf{d}, middle panel) in the ordered phase, and compares well to our
theoretical calculations (Fig.~\ref{fig:path}\textbf{d}, upper panel).
In Fig.~\ref{fig:path}{\bf e} we show the spectral weight along the
diagonal cut of the Brillouin zone. We observe the presence of a 
sharp kink in the dispersion (\emphasize{waterfall}) that was
also recently reported in experiments \cite{moritz_waterfall_ncco}. 

In Fig.~\ref{fig:fsdop} we show the doping evolution of the Fermi surface
obtained by theoretical calculations (lower panels). We find that upon doping,
the Fermi surface moves from the square-like magnetic Fermi-surface
towards the Fermi arcs of the paramagnetic Fermi surface (panels d-g). 
This is explained by the closing of the 
pseudo-gap at $\bold k=(\pi/2,\pi/2)$.
The agreement with experiments (panels a-c) \cite{armitage_fermi_prl} is very satisfactory.

%%%%%%%%%%%%%%%%%%%%%%%%%%%%%%%%%%%%%%%%%%%%%%%%%%%%%%%%%%%%%%%%
%%%%%%%%%%%%%%%%%%%%%%%%%%%%%%%%%%%%%%%%%%%%%%%%%%%%%%%%%%%%%%%%
%%%%%%%%%%%%%%%%%%%%%%%%%%%%%%%%%%%%%%%%%%%%%%%%%%%%%%%%%%%%%%%%
%%%%%%%%%%%%%%%%%%%%%%%%%%%%%%%%%%%%%%%%%%%%%%%%%%%%%%%%%%%%%%%%
%%%%%%%%%%%%%%%%%%%%%%%%%%%%%%%%%%%%%%%%%%%%%%%%%%%%%%%%%%%%%%%%
%%%%%%%%%%%%%%%%%%%%%%%%%%%%%%%%%%%%%%%%%%%%%%%%%%%%%%%%%%%%%%%%

\section{Photoemission of LSCO}

In this section we focus on the spectral functions of LSCO. 
In particular, we focus on the 6-band theory, which includes \dz and \pz orbitals,
which are expected to play a role in LSCO, due to the presence of apical oxygens.
Note that the apical oxygens are absent in NCCO.

\begin{figure}
\begin{center}
\includegraphics[width=\figwidth]{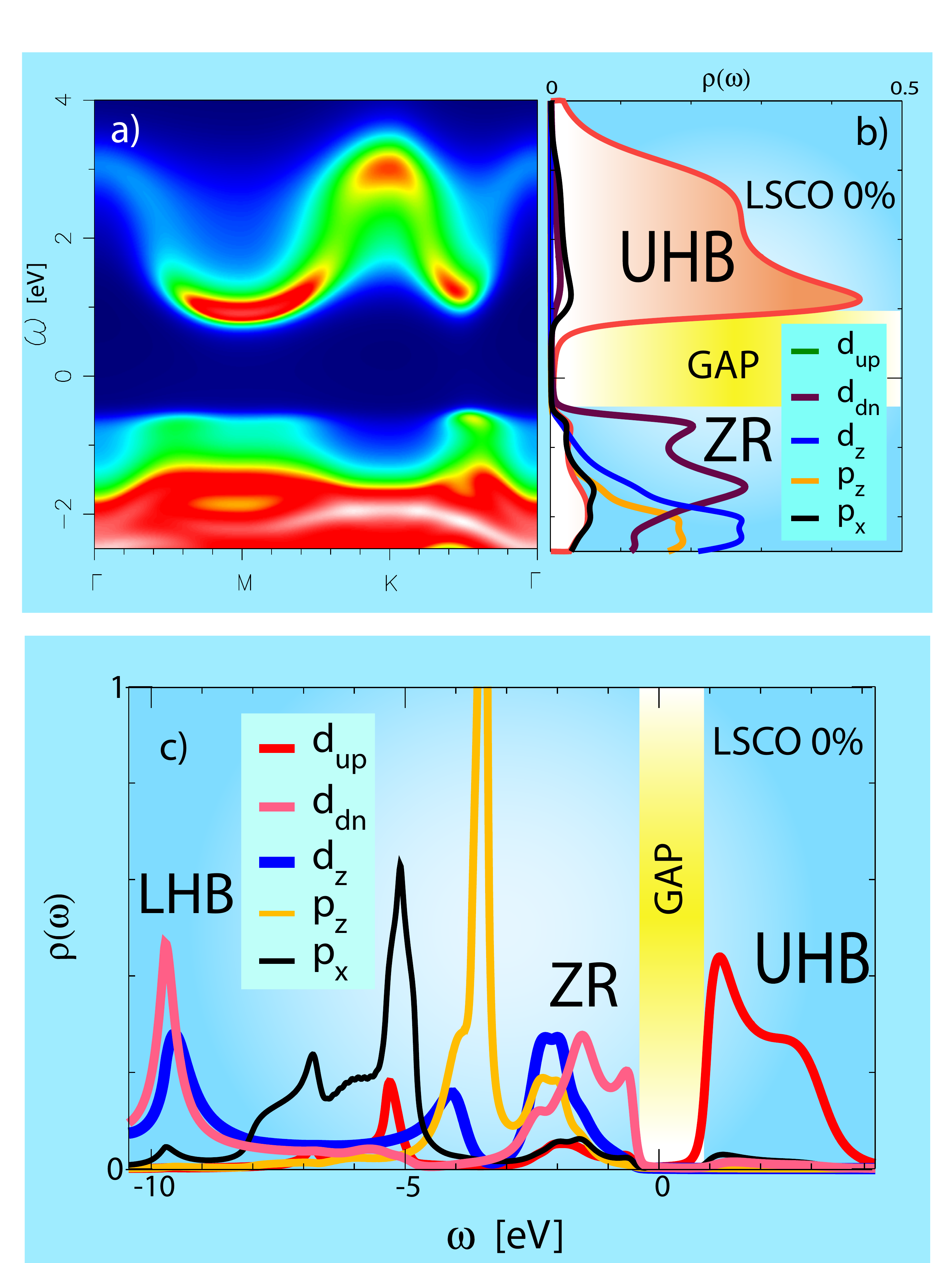}
\caption{
(Colors online) 
a) Frequency dependent spectral weight A(k,w) obtained by LDA+DMFT of a 6-band
model description of the parent compound of LSCO. 
b) Partial density of states of the \dx, \dz, \px and \pz orbitals.
We observe a direct gap of $1.8$eV in LSCO. Notice that the spectral weight is very incoherent close to the Fermi energy
in the lower band. 
c) Partial density of states on a larger energy scales. 
The lower Hubbard band (LHB) is located at a very low energy $-10$eV, and the upper Hubbard band (UHB) is also shown.
The \dz and \pz orbitals have a strong weight between $-4$eV and $-1$eV.
}
\label{fig:6band_parent}
\end{center}
\end{figure}

In Fig.~\ref{fig:6band_parent}{\bf a} we show the momentum resolved spectral function of LSCO obtained
by DMFT. We observe a direct gap $\approx 1.8$eV, which is larger than the gap in NCCO, showing
that LSCO is more correlated than NCCO. 
The partial density of states (Fig.~\ref{fig:6band_parent}{\bf b}) shows two
dispersive features, the upper Hubbard band (UHB) and a the band below the Fermi level, the Zhang-Rice singlet (ZR). The latter is
an admixture of oxygen and copper characters. 
The Zhang-Rice singlet is more incoherent in LSCO than in NCCO (see Fig.~\ref{fig:DOS}{\bf a}). 

Fig.~\ref{fig:6band_parent}{\bf b} is a blow up of 
Fig.~\ref{fig:6band_parent}{\bf c}, that displays the integrated spectrum on a larger energy scale. 
The lower Hubbard band (LHB) is separated from the upper Hubbard band (UHB) by an energy scale of the order of $U_d$.
It is worth noting that the \pz and \dz orbitals have a strong weight 
between $-4$eV and $-1$eV, and the in-plane oxygens are located at $-5$eV.
Hence, the additional orbitals \pz and \dz hybridize with the Zhang-Rice singlet,
and change the theoretical description of LSCO for energies larger 
than 1eV.

\begin{figure}
\begin{center}
\includegraphics[width=\figwidth]{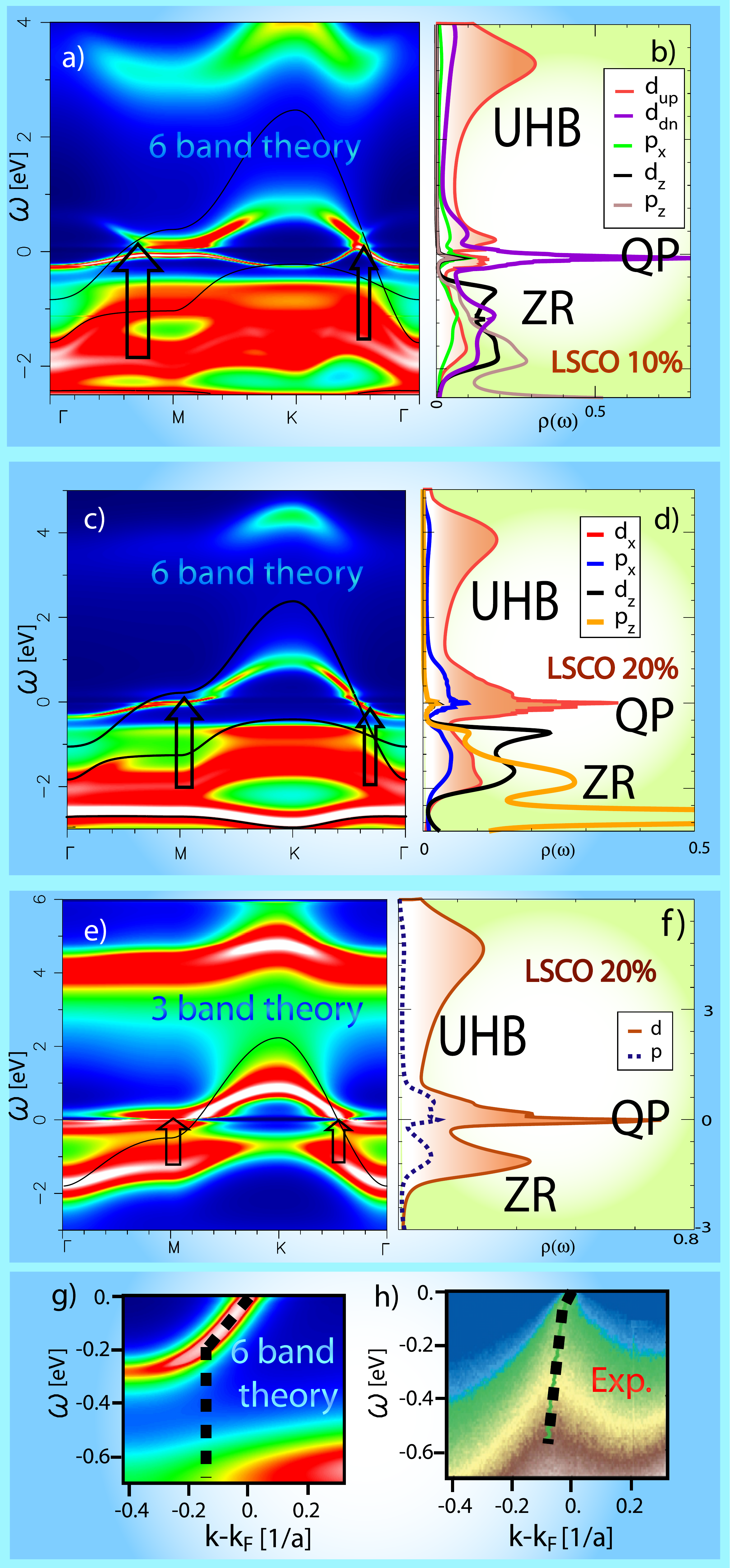}
\caption{
(Colors online) a),c) Frequency dependent spectral weight A(k,w) obtained by LDA+DMFT of a 6-band model description of doped LSCO. 
b) and d) Partial density of states of the \dx, \dz, \px and \pz orbitals. a) and b) are obtained in the ordered state.
e) and f) are obtained within a 3-band description of doped LSCO and are shown for comparison.
The optical transitions occur at different energies for the 3-band and 6-band theories (see vertical arrows in c and e).
Also note that the weight of the \dz and \pz orbitals in b) and d) close to the Fermi level
is consequent, which justifies that a 6-band description is necessary. The solid lines in a),c) and e) are the rigid LDA bands.
g) A(k,w) along the nodal cut of the Brillouin Zone for the paramagnetic 6-band theory at $10\%$ doping, compared side by side with 
h) experimental data of Ref.~\onlinecite{valla_kink}. The dashed line are guide to the eyes to illustrate the sharp kink (\emphasize{waterfall})
in the dispersion.
}
\label{fig:6band_doped}
\end{center}
\end{figure}

In Fig.~\ref{fig:6band_doped}{\bf a} and \ref{fig:6band_doped}{\bf c} we focus on the momentum resolved spectral
functions of doped LSCO. Figures \ref{fig:6band_doped}{\bf b} and \ref{fig:6band_doped}{\bf d} are the corresponding integrated 
spectral functions. At $10\%$ doping (Fig.~\ref{fig:6band_doped}{\bf b}), the Zhang-Rice singlet has an incoherent contribution (ZR)
and a coherent part - the quasi-particle peak (QP). The coherent part (QP) is the narrow band below the Fermi
level in Fig.~\ref{fig:6band_doped}{\bf a}. 
The vertical arrows in Fig.~\ref{fig:6band_doped}{\bf a} and \ref{fig:6band_doped}{\bf c} highlight the location of the direct transitions from occupied
states to unoccupied states. 
These transitions are important for the optical conductivity (which we discuss in the next section).

For comparison, we also show the theoretical description of LSCO without the apical oxygens in Fig.~\ref{fig:6band_doped}{\bf e-f}. 
The main difference between the 3-band and the 6-band descriptions, is 
that the incoherent part of ZR is narrower in the 3-band theory.
The vertical transitions, marked with vertical arrows in 
Fig.~\ref{fig:6band_doped}{\bf c,e}, 
highlight large contributions to the optical conductivity,
and one can see that the optical transitions occur at different energies in the two
models.

Fig.~\ref{fig:6band_doped}{\bf g} is the momentum resolved spectral function around a nodal cut of the Brillouin zone. We observe the presence
of a very sharp kink in the dispersion (\emphasize{waterfall}), 
in agreement with experimental data  
of Ref.~\onlinecite{valla_kink} reproduced in Fig.~\ref{fig:6band_doped}{\bf h}.

\begin{figure}
\begin{center}
\includegraphics[width=\figwidth]{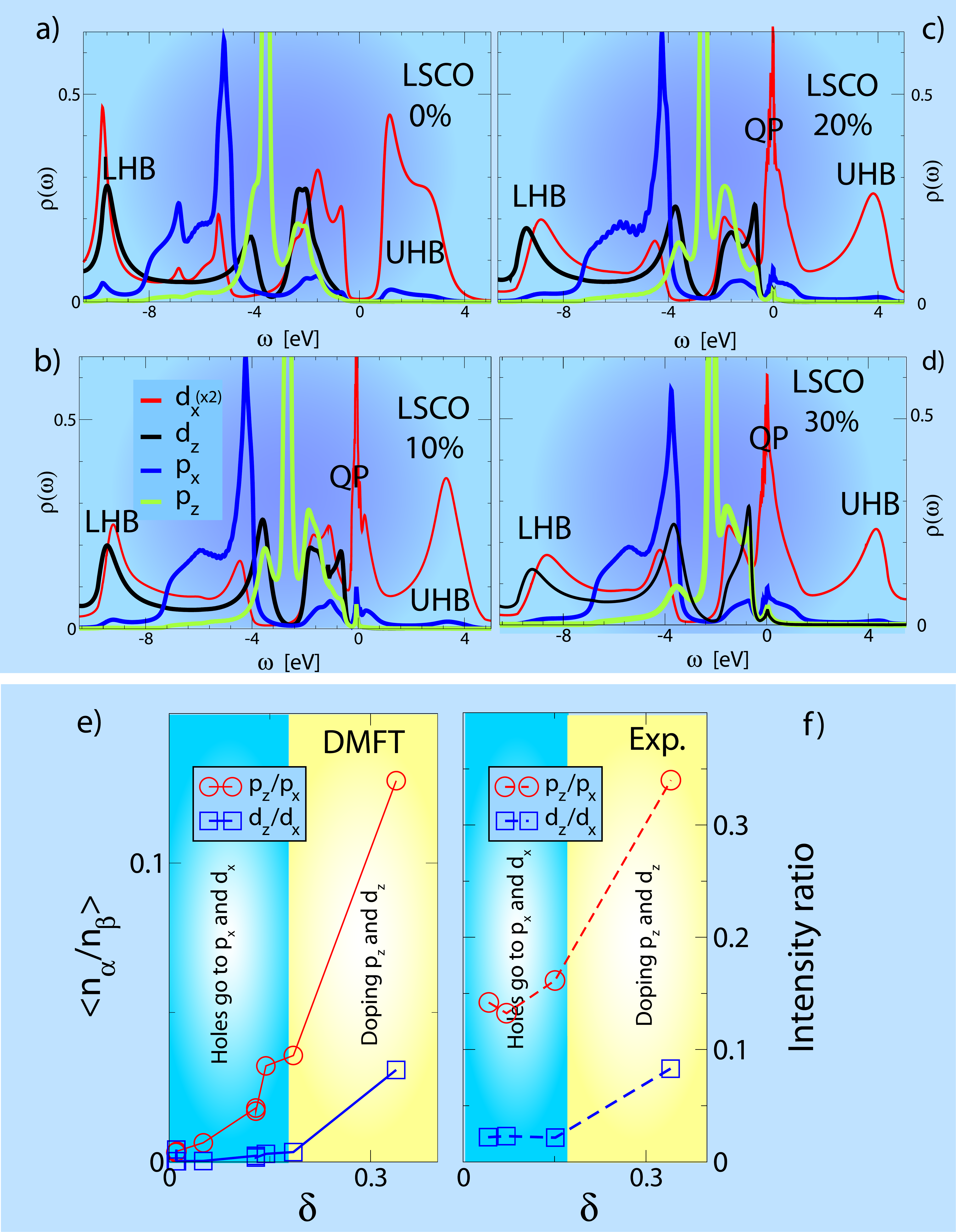}
\caption{
(Colors online) 
Spectral functions for the 6-band theory of LSCO in a) the parent compound and at b) $10\%$ and c) $20\%$ doping and d) $30\%$ doping.
The \dx, \dz, \px and \pz components are shown. Note that the lower Hubbard band (LHB) at $-9$eV is both of \dx and \dz character.
At finite doping (b) and (c), there is a quasi-particle peak (QP) close to the Fermi energy. Note that the \dz and \pz weight gets
closer to the Fermi level for higher doping (c). These latter orbitals will therefore be important to describe high doping and/or high energy spectroscopy.
e) Ratio of the hole occupancy of the \dz and \dx orbitals (squares) and of the \pz and \px orbitals (circles) obtained by DMFT, and f)
experimental data are also shown for comparison \cite{apical_exp}.
}
\label{fig:6band_dos_doped}
\end{center}
\end{figure}

Fig.~\ref{fig:6band_dos_doped}{\bf a-d} displays the orbital resolved spectral functions, obtained within a 6-band theory, for various dopings on 
a wide energy scale. Upon doping (Fig.~\ref{fig:6band_dos_doped}{\bf b-d}) we observe the presence of the quasi-particle
peak, and the upper Hubbard band (UHB) smears out. Additionally, in the doped compound
the \dz and \pz orbitals have a larger weight close to the Fermi level. 

Fig.~\ref{fig:6band_dos_doped}{\bf e} show the relative occupation (in hole notation)
of the \dz and \pz orbitals upon doping. 
Our results are in agreement with Ref. \cite{more_ref21} and we find that even after
inclusion of the apical oxygens, there is no 
saturation of the occupancy observed around doping 0.2 in
the x-ray absorption spectroscopy (XAS)
experiments of Ref. \cite{xray_absorption}. We notice however
that LDA+DMFT does capture the evolution of the {\it  ratio
of the occupancies} of apical and planar oxygens:
For doping $\delta < 20\%$, the holes go mainly to the
\dx and \px orbitals, and for larger doping $\delta > 20\%$, the holes start to fill the \dz and
\pz. 
In Fig.~\ref{fig:6band_dos_doped}{\bf f} we report the experimental data of Ref. \cite{apical_exp} for side-by-side comparison
with DMFT calculations. Our calculations reproduce the rapid increase of the occupancy of $p_z$ around doping 0.2.

We note that modeling XAS more accurately may require
downfolding the LDA bands over a much larger energy range to include more
orbitals, or the inclusion of the doping dependence modifications
of the apical oxygen distance to the copper oxide layer pointed
out in Ref \cite{more_ref16}. 
A more accurate modeling of the XAS, including the core hole
potential, as it was done for the core level photoemission in Ref
\cite{cornaglia}, might also be necessary.

%%%%%%%%%%%%%%%%%%%%%%%%%%%%%%%%%%%%%%%%%%%%%%%%%%%%%%%%%%%%%%%%
%%%%%%%%%%%%%%%%%%%%%%%%%%%%%%%%%%%%%%%%%%%%%%%%%%%%%%%%%%%%%%%%
%%%%%%%%%%%%%%%%%%%%%%%%%%%%%%%%%%%%%%%%%%%%%%%%%%%%%%%%%%%%%%%%
%%%%%%%%%%%%%%%%%%%%%%%%%%%%%%%%%%%%%%%%%%%%%%%%%%%%%%%%%%%%%%%%
%%%%%%%%%%%%%%%%%%%%%%%%%%%%%%%%%%%%%%%%%%%%%%%%%%%%%%%%%%%%%%%%
%%%%%%%%%%%%%%%%%%%%%%%%%%%%%%%%%%%%%%%%%%%%%%%%%%%%%%%%%%%%%%%%
%%%%%%%%%%%%%%%%%%%%%%%%%%%%%%%%%%%%%%%%%%%%%%%%%%%%%%%%%%%%%%%%
%%%%%%%%%%%%%%%%%%%%%%%%%%%%%%%%%%%%%%%%%%%%%%%%%%%%%%%%%%%%%%%%
%%%%%%%%%%%%%%%%%%%%%%%%%%%%%%%%%%%%%%%%%%%%%%%%%%%%%%%%%%%%%%%%
%%%%%%%%%%%%%%%%%%%%%%%%%%%%%%%%%%%%%%%%%%%%%%%%%%%%%%%%%%%%%%%%
%%%%%%%%%%%%%%%%%%%%%%%%%%%%%%%%%%%%%%%%%%%%%%%%%%%%%%%%%%%%%%%%
%%%%%%%%%%%%%%%%%%%%%%%%%%%%%%%%%%%%%%%%%%%%%%%%%%%%%%%%%%%%%%%%
%%%%%%%%%%%%%%%%%%%%%%%%%%%%%%%%%%%%%%%%%%%%%%%%%%%%%%%%%%%%%%%%
%%%%%%%%%%%%%%%%%%%%%%%%%%%%%%%%%%%%%%%%%%%%%%%%%%%%%%%%%%%%%%%%
%%%%%%%%%%%%%%%%%%%%%%%%%%%%%%%%%%%%%%%%%%%%%%%%%%%%%%%%%%%%%%%%
%%%%%%%%%%%%%%%%%%%%%%%%%%%%%%%%%%%%%%%%%%%%%%%%%%%%%%%%%%%%%%%%
%%%%%%%%%%%%%%%%%%%%%%%%%%%%%%%%%%%%%%%%%%%%%%%%%%%%%%%%%%%%%%%%
%%%%%%%%%%%%%%%%%%%%%%%%%%%%%%%%%%%%%%%%%%%%%%%%%%%%%%%%%%%%%%%%

\clearpage

\section{Optical properties of LSCO and NCCO}

\begin{figure}
\begin{center}
\includegraphics[width=\smallwidth]{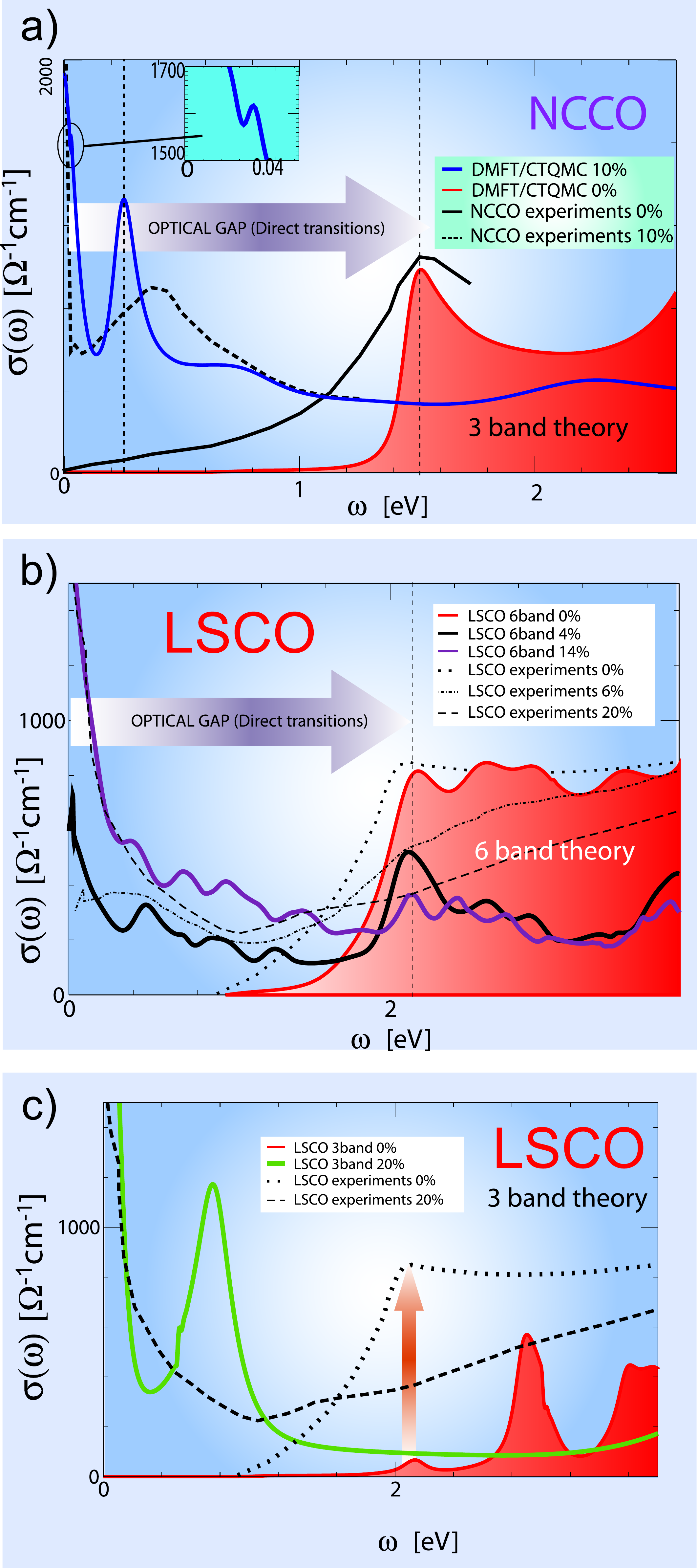}
\caption{
(Colors online) 
a) Theoretical optical conductivity of NCCO in $\Omega^{-1}cm^{-1}$ 
at integer-filling (red line). We find that NCCO has an optical gap of about 1.5eV,
which is larger than the direct gap $\approx 1.2$eV. We also show data for doped NCCO (blue line).
Note that for doping smaller than $20\%$, 
we observe a small peak at a small energy $\approx 0.035eV$ (see the inset).
We also observe a second peak at larger energy
scale $\approx 0.2 eV$, which corresponds to optical transitions within the Zhang-Rice singlet (see Fig.~\ref{fig:DOS}{\bf c}). 
For comparison we also show the infrared optics of Refs \cite{infrared_optics,infrared_optics_2} (dashed black line).
b) Optical conductivity of the 6-band theory of LSCO for the parent compound (red line) and doped LSCO.
For comparison we also show experimental data of Ref.~\onlinecite{uchidalsco} (short and long dashed lines).
c) Optical conductivity of the 3-band theory of LSCO in the parent compound (red line) and doped LSCO (yellow
line), and experimental data of Ref.~\onlinecite{uchidalsco} (short and long dashed lines).
The vertical arrow in panel (c) emphasizes the disagreement between theory and experiments.
Note also that at finite doping there is a peak at $0.8$eV in the optical conductivity of the 3-band
theory, which is absent from experimental data. This peak is related to the optical transitions shown in
Fig.~\ref{fig:6band_doped}{\bf e} (see the vertical arrows).  All calculations were done in the ordered state.
}
\label{fig:optic}
\end{center}
\end{figure}

We now turn to the optical conductivity.  It was previously computed for the 3-band model of LSCO
\cite{our_previous_paper_lsco} and 
we now generalize the results for the 6-band description of LSCO
and we also compute it for NCCO (see Fig.~\ref{fig:optic}).

The optical conductivity in LDA+DMFT is given by:
\begin{multline}
 \sigma'(\omega) = \frac{1}{N_k} \sum\limits_{\sigma \vk}{\frac{\pi e^2}{\hbar c }  \int{ dx \frac{  f(x-\omega)-f(x)}{\omega}}}
                  \\ \times \textrm{Tr}\big( \hat{\bold{\rho}}_{\vk\sigma} (x - \omega)  \bold{v}_\vk  \hat{\bold{\rho}}_{\vk\sigma}(x) \bold{v}_\vk \big)
\end{multline}
Where $c$ is the interlayer distance, and the density matrix
$\bold{\hat{\rho}}$ is defined by
\begin{equation}
  \hat{\bold{\rho}}_{\vk\sigma}(x)=\frac{1}{2\pi i}
\left(\bold{G}^\dagger_{\vk\sigma}(x)-\bold{G}_{\vk\sigma}(x)\right)
\end{equation}
The bare vertex for a multiple orbital problem $v_k^{\alpha,\beta}=\frac{dH_k^{\alpha,\beta}}{dk_x}+i\left( q_x^\alpha-q_x^\beta  \right) H_k^{\alpha,\beta}$ 
is obtained following the steps of Ref~\onlinecite{tomczak}. The Peierl phase $i\left( q_x^\alpha-q_x^\beta  \right) H_k^{\alpha,\beta}$ plays an important
role in particular for the ordered state (as discussed in 
Ref~\onlinecite{tomczak}, if this phase is not considered the optical conductivity depends on 
any artificial folding of the Brillouin zone).

In Fig.~\ref{fig:optic}{\bf a} we show 
the theoretical optical conductivity of NCCO 
at integer filling (red curve) and at 10 percent doping (blue curve).
The undoped compound has a sharp onset at an energy of the order of 1.5 eV
which we interpret as the direct gap (slightly larger than the charge transfer gap in Fig.~\ref{fig:DOS}\textbf{b}.

Doping introduces several new features (blue line in Fig.~\ref{fig:optic}\textbf{a}). The $1.5 eV$ optical peak disappears and
the weight is transferred to lower energy in the form of a Drude peak and a
mid infrared peak at $\omega \approx 0.2eV$.
The optical conductivity also displays a peak in the magnetic solution
at a much lower frequency $\omega \approx 0.035eV$ (see left inset of Fig.~\ref{fig:optic}\textbf{a}).

\begin{figure}
\begin{center}
\includegraphics[width=\figwidth]{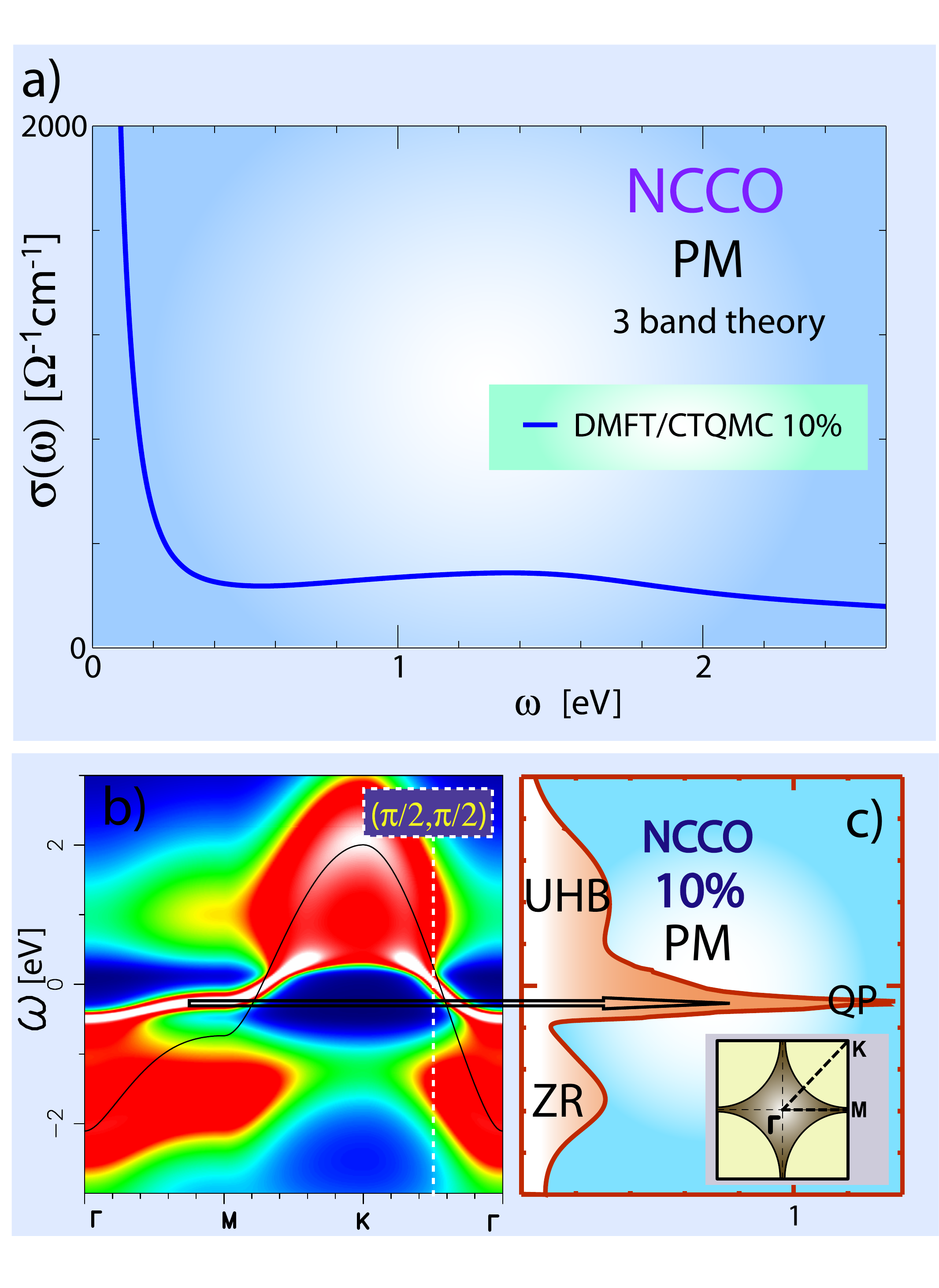}
\caption{
(Colors online)
a) Theoretical optical conductivity of paramagnetic NCCO at $10\%$ doping is shown.
In the paramagnetic state of NCCO, there are no peaks at $0.035$eV and $0.2$eV which
are observed in the magnetic state of NCCO at same doping (see Fig.\ref{fig:optic}{\bf a}).
b) We show the k dependent spectral weight $A(k,\omega)$ for NCCO at $10\%$ doping.
c) We show the corresponding total density of states. In the paramagnetic state of NCCO there is
no splitting of the quasi-particle peak QP (see Fig.~\ref{fig:DOS}{\bf d}).
}
\label{fig:ncco_pm_sdw_comparison}
\end{center}
\end{figure}

Below 0.5 eV, vertical transitions are among
the quasi-particle bands of the magnetic DMFT band structure.
This involves a continuum of k points, but it is likely to be
controlled by saddle points in the reciprocal space. One saddle is at $M=(\pi,0)$ and
transitions close to that point, indicated by a vertical arrow in Fig.~\ref{fig:DOS}\textbf{c},
give rise to the peak in the optical conductivity at 0.2 eV.
Transitions close to the midpoint between $K=(\pi,\pi)$ and
$\Gamma=(0,0)$, give rise to the small peak at 0.035 eV in the optical conductivity.
Both peaks are characteristics of the quasi-particle band structure in the magnetic
state of NCCO, and these features are not present in the absence of magnetic order,
as shown in Fig.~\ref{fig:ncco_pm_sdw_comparison}{\bf a}. The absence of the 
peak at 0.035eV and 0.2eV in the optical conductivity is explained by the absence
of vertical transition in the k dependent spectral weight $A(k,\omega)$ (see Fig.~\ref{fig:ncco_pm_sdw_comparison}{\bf b}),
which are present in the ordered state and highlighted by the vertical arrow of Fig.~\ref{fig:DOS}{\bf c}.

The quasi-particle peak in the ordered state is split into two narrow bands, 
one above and one below the Fermi level (see Fig.~\ref{fig:DOS}{\bf d}),
while the splitting is absent in the paramagnetic state 
(see Fig.~\ref{fig:ncco_pm_sdw_comparison}{\bf c}).
This is due to the absence of the pseudo-gap around $(\pi/2,\pi/2)$ in the paramagnetic state (see Fig.~\ref{fig:ncco_pm_sdw_comparison}{\bf b}).
As a consequence, the paramagnetic Drude peak is featureless, while in the ordered state 
we observe several low energy peaks in the optical conductivity (at 0.035 eV and 0.2 eV).

The agreement between DMFT and experimental data \cite{infrared_optics,infrared_optics_2}
(dashed line of Fig.~\ref{fig:optic}{\bf a}) is qualitative and our theory
connects the peak in the experimental optical conductivity at 0.4 eV
with magnetism, in agreement with Ref. \cite{spin_waves_ncco}. Additionally our study allow us 
to connect this peak with the spectral weight below the Fermi energy at the $M=(\pi,0)$ point,
which is present in both the paramagnetic and the ordered state.

We note finally that the peak in the optical conductivity at smaller energy 0.035eV
is observed at this energy in experiments (see Fig. 2 of Ref.~\onlinecite{tokura_small_peak}),
which is only present within the ordered state and disappears at higher temperature in the paramagnet,
which suggest that this feature is not connected to phonons.
Our calculation connects this peak to the presence of magnetism and to the pseudo-gap
around $(\pi/2,\pi/2)$ (see Fig.~\ref{fig:DOS}\textbf{c}). 

We now turn to the description of the optical conductivity of LSCO.
In Fig.~\ref{fig:optic}{\bf b} we show the optical conductivity of the 6-band description of LSCO
in the parent compound (red curve), at $4\%$ (black curve) and $16\%$ doping (blue curve).
For comparison we also show experimental data of Ref.~\onlinecite{uchidalsco}
at $0\%$ doping (dotted line), $6\%$ (short dashed) and $20\%$ (long dashed).
Note that there is a quantitative agreement between the theory and experiments for frequencies
smaller $\omega<2$eV. At larger frequencies $\omega>2$eV, the optical
conductivity in experiments is larger than the theoretical one,
 which is due to optical transition
from additional orbitals that are not present within our calculations, and which contributes
at high energies.

For comparison, we now also consider the three band theory of LSCO. 
In Fig.~\ref{fig:optic}{\bf c} we show the optical conductivity of the magnetic state 
of the parent compound (red curve) and of the doped LSCO at $20\%$ doping (yellow curve).
We also show the experimental data of Ref.~\onlinecite{uchidalsco}
at $0\%$ doping (dotted line) and $20\%$ doping (dashed line). 
The vertical arrow emphasizes the disagreement between theory and experiments.
The strong differences between the 3-band and 6-band calculations for
the optical conductivity of the parent compound is related to the strong weight of the
\pz and \dz orbitals below the Fermi level, which are mainly occupied and contribute
significantly to the optical conductivity. This weight is obviously absent in the three band theory.
Note also that at finite doping there is a peak at $0.8$eV in the optical conductivity of the 3-band
theory, which is absent in experimental data. This peak is related to the optical transitions shown in
Fig.~\ref{fig:6band_doped}{\bf e} by the vertical arrows. This artifact of the 3-band calculations is cured by
introducing the \dz and \pz orbitals. 

To quantify the rate of the redistribution of optical spectral
weight, we computed the effective electron number per Cu atom
defined by
\begin{equation}
N^{\Lambda}_{eff}=\frac{2 m_e V}{\hbar \pi e^2}
\int^{\Lambda}_{0}{\sigma'(\omega) d\omega}, 
\label{Sweight}
\end{equation}
where $m_e$ is the free electron mass, and $V$ is the cell volume
containing one formula unit. $N_{eff}$ is proportional to the
number of electrons involved in the optical excitations up to the
cutoff $\Lambda$.

\begin{figure}
\begin{center}
\includegraphics[width=\figwidth]{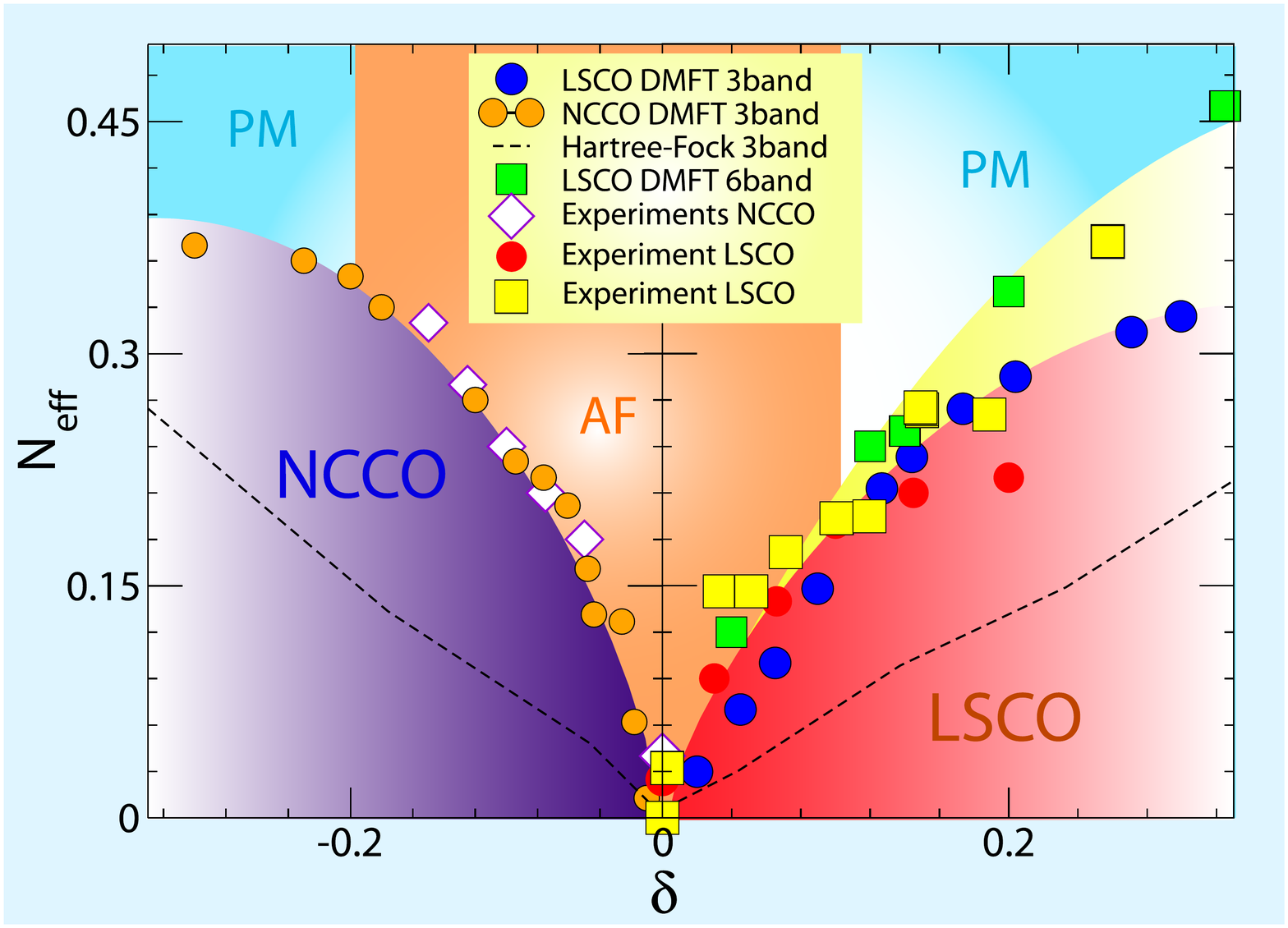}
\caption{
(Colors online)
We show the dimensionless integrated optical conductivity $N_{eff}$ for the 3-band study of LDA+DMFT done on LSCO \cite{our_previous_paper_lsco} and NCCO, obtained
with a cutoff $\Lambda=1.2$ ($\Lambda=1.5$) for NCCO (LSCO).
Experimental data for LSCO (red circles \cite{experimental_Nf} and yellow squares \cite{lupilscooptic}), and NCCO \cite{tokura_optics_ncco} (open diamonds) are shown.
The increase of $N_{eff}$ is similar for both compounds.
The dashed line indicates the results obtained by the Hartree Fock approximation.
We also show for comparison results obtained within the 6-band theory (green squares).
The agreement between theory and experiments is quantitative.
}
\label{fig:Neff}
\end{center}
\end{figure}

Our results for $N_{eff}$ are displayed in
Fig.~\ref{fig:Neff} and compared to experimental data taken from
Ref.~\cite{experimental_Nf},~\cite{tokura_optics_ncco} and ~\cite{lupilscooptic}.
Notice a favorable agreement between the theory and experiment, for
which the use of the realistic electronic structure is essential.
Moreover, we also show in Fig.~\ref{fig:Neff} the static mean-field results. We emphasize  
that the static mean field is not able to reproduce qualitatively the experimental trend.
This highlights the importance
of the dynamical fluctuations not taken into account in Hartree Fock.
The trend of $N_{eff}$ is qualitatively similar for both LSCO and NCCO, and therefore
does not show that LSCO and NCCO are qualitatively different: the former being a Mott 
and the latter being a Slater insulator. 

\begin{figure}
\begin{center}
\includegraphics[width=\figwidth]{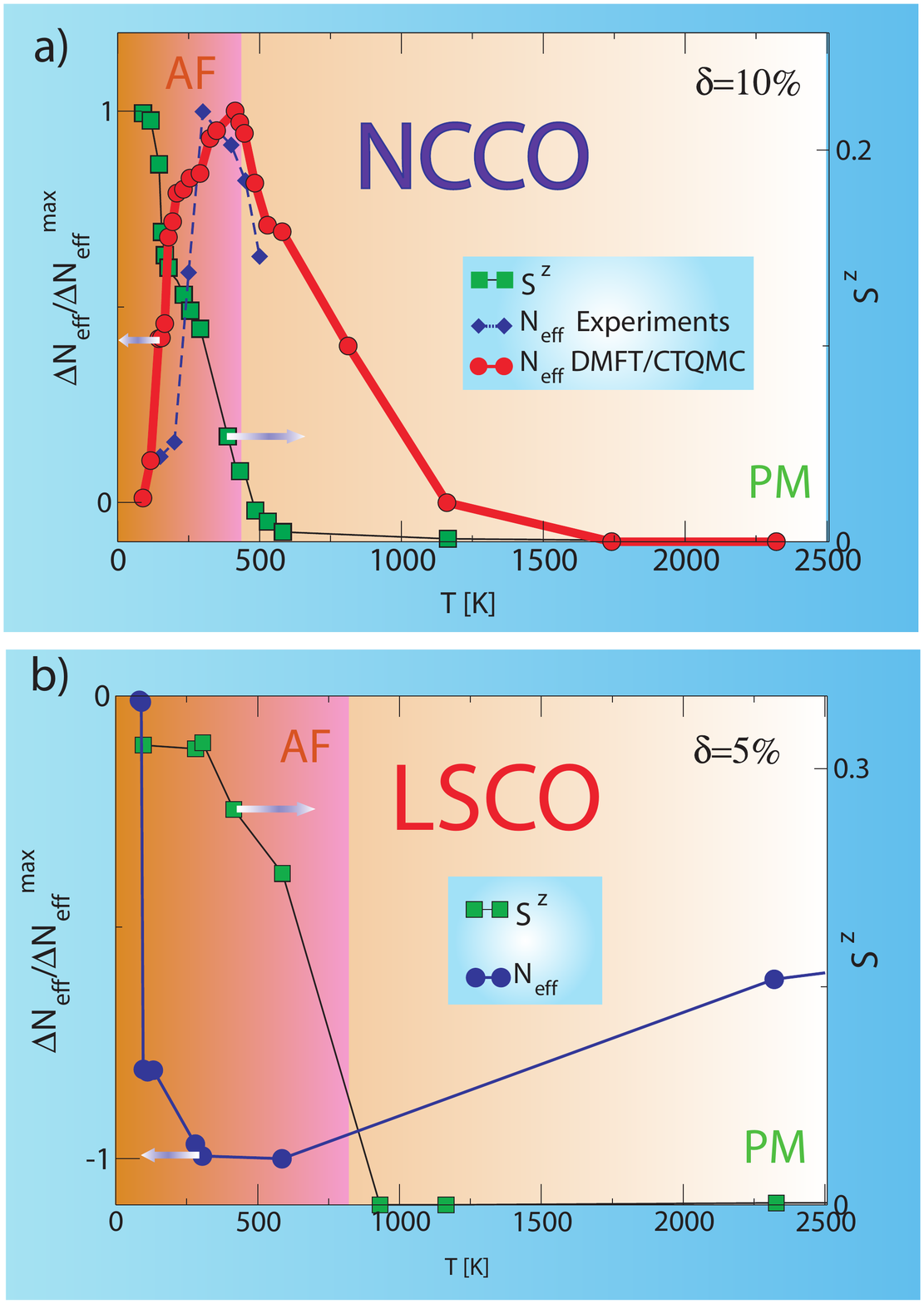}
\caption{
(Colors online)
a) 3-band theoretical normalized variation of $N_{eff}$, $\Delta N_{eff} = N_{eff}(T)-N_{eff}(T=89^\circ K)$, at $10\%$ electron doping for NCCO (red circles, left scale).
$N_{eff}$ is reaching a maximum value when the magnetization (orange squares, right scale) is destroyed by the thermal fluctuations.
The decrease of $N_{eff}$ at low temperature can be explained by the opening of a pseudo-gap in the ordered phase.
The data were obtained by single site DMFT calculations in the ordered phase. 
The dashed line corresponds to experiments (see Fig.~7{\bf c} of Ref \onlinecite{tokura_optics_ncco}),
where they measured $N_{eff}(\Lambda=0.03eV)$ (contribution due to the Drude peak) 
and $N_{eff}(\Lambda=0.3eV)-N_{eff}(\Lambda=0.2eV)$ (contribution due to the pseudogap). The dashed
line corresponds to the sum of these two contributions.
b) 3-band single site DMFT of the ordered phase of LSCO is shown for $5\%$ doping. Note that the trend of $N_{eff}$ is opposite 
between LSCO AND NCCO, which is a signature that NCCO is a Slater insulator and LSCO a Mott insulator.
}
\label{fig:neff_temp}
\end{center}
\end{figure}

We also compute the temperature dependence of $N_{eff}$ for NCCO and LSCO at a fixed density.
In Fig.~\ref{fig:neff_temp}{\bf a} we show the theoretical results for NCCO. 
We plot in the same figure  
the temperature dependence of $N_{eff}$ (left scale) and the temperature
dependence of the magnetic moment (right scale).
We find that $N_{eff}$ is reaching a maximum value
when the magnetization is suppressed by thermal fluctuations at $\approx 400$K.
There is actually a variation of $N_{eff}$ inside
the ordered phase, that can be explained by the
destruction of the magnetic pseudo-gap.
Once the magnetization is quenched by the temperature,
heating the system further reduces $N_{eff}$.
The qualitative agreement with experimental 
data extracted from Ref.~\cite{tokura_optics_ncco} is very encouraging.

The temperature dependence of $N_{eff}$ for LSCO is shown in Fig.~\ref{fig:neff_temp}{\bf b}.
We emphasize that the temperature dependence of $N_{eff}$ shows an opposite trend for LSCO.
When doping the parent compound, there is a {\it decrease} of $N_{eff}$ in LSCO, whereas there
is an {\it increase} of $N_{eff}$ in NCCO.  
Therefore the temperature dependence of $N_{eff}$ clearly shows a distinct behavior for
a Slater and a Mott insulator.
Note that the same general trend of the temperature dependence of $N_{eff}$ is observed in
experiments for LSCO, but for a larger doping than the one considered here 
(optical data for $13\%$ doping in Fig. 6 of Ref. \cite{basov_neff}).

\begin{figure}
\begin{center}
\includegraphics[width=\figwidth]{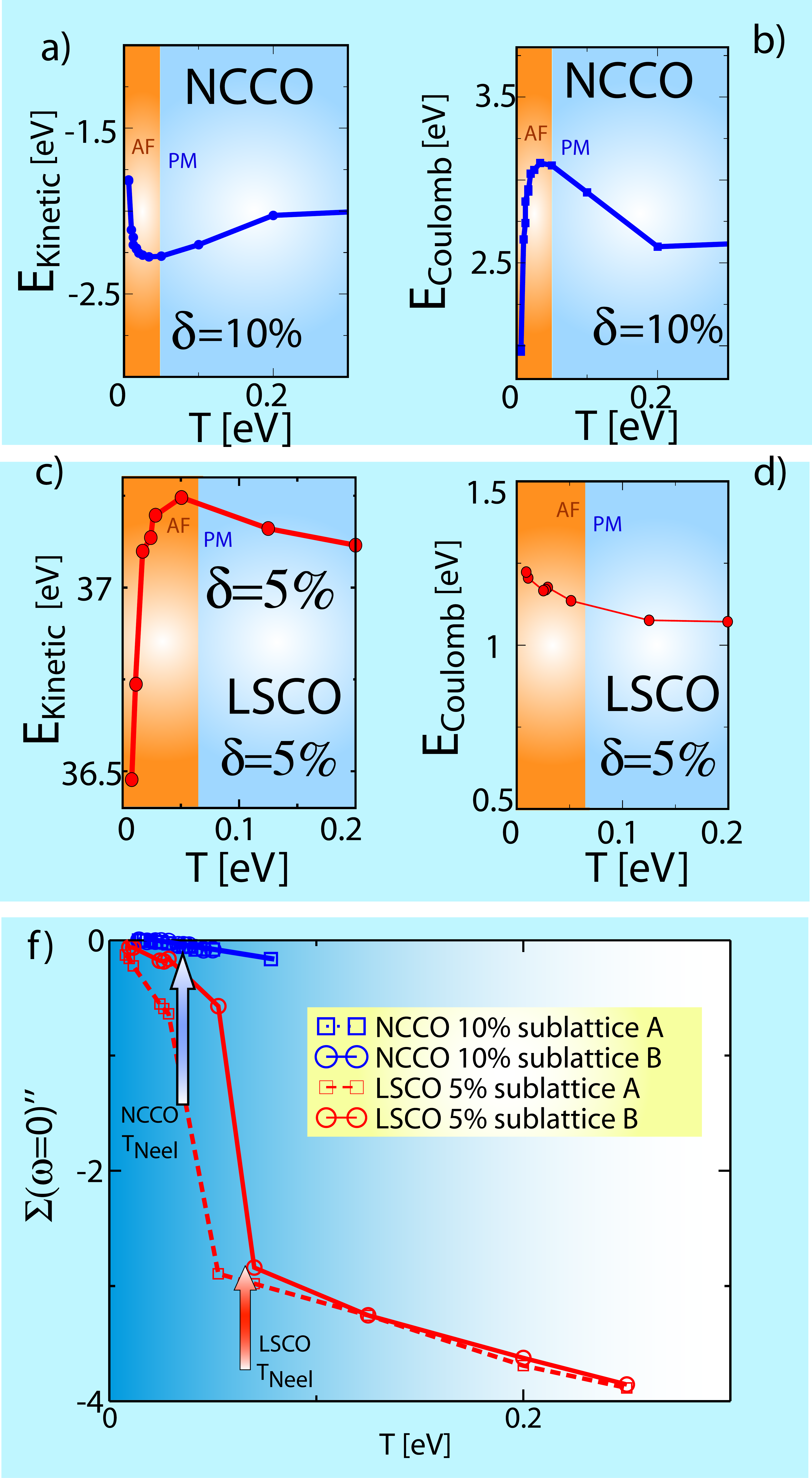}
\caption{
(Colors online) 3-band theoretical temperature dependence of the a) Kinetic energy $\mathcal{H}_t$ (equation \ref{eq:3band_hub1})
and b) Coulomb enery $\mathcal{H}_U$ (equation \ref{eq:3band_hub2}) of NCCO at $10\%$ doping, and c) Kinetic and d) Coulomb
energy of LSCO at $5\%$ doping. The red area highlights the temperature region where the solution is magnetic (AF), and the solution is paramagnetic (PM) in
the blue area.
a)-b) are showing that there is a kinetic energy optimization when LSCO becomes an antiferromagnet, which is proper to the
Mott insulator, and c)-d) show that NCCO is a typical Slater insulator, which optimizes the Coulomb (local onsite repulsion) energy when it becomes
an antiferromagnet, at the expense of a worse Kinetic energy.
This is consistent with the theoretical $N_{eff}$ shown in Fig.~\ref{fig:neff_temp}.
f) We show the imaginary part of the self energy at zero frequency $Im\Sigma(\omega=0)$ in function of the temperature.
The scattering rate of LSCO $\lambda = -Im\Sigma(\omega=0) $ (red lines) is strongly reduced for $T<T_{Neel}$ ($T_{Neel}$ is
highlighted by the red vertical arrow),
which is consistent with the kinetic energy reduction observed for LSCO at low temperature (Fig.~\ref{fig:kinetic_energy}{\bf c}).
NCCO (blue lines) is showing a small scattering rate weakly dependent on the temperature, which
is consistent with NCCO being more metallic than LSCO.
All calculations were obtained by CTQMC in the ordered state.
}
\label{fig:kinetic_energy}
\end{center}
\end{figure}

The trend of $N_{eff}$ can be understood in a simple picture.
In a Slater picture, the onset of antiferromagnetism reduces the
Coulomb correlations (double occupancy) at the expense of the kinetic
energy. The opposite is true in a charge transfer insulator. Consequently, in a
Slater insulator the kinetic energy becomes less negative as the
temperature decreases while the opposite happens in a charge transfer insulator.
The kinetic energy as a function of temperature is readily available
in the theory and is displayed in Fig.~\ref{fig:kinetic_energy}\textbf{a} and Fig.~\ref{fig:kinetic_energy}\textbf{c} and the 
Coulomb energy is displayed in Fig.~\ref{fig:kinetic_energy}\textbf{b} and Fig.~\ref{fig:kinetic_energy}\textbf{d}.

In Fig.~\ref{fig:kinetic_energy}\textbf{f} we display the imaginary part of 
the self energy at zero frequency, $Im\Sigma(\omega=0)$,
as a function of the temperature, which is proportional 
to the scattering rate $\lambda = -Im\Sigma(\omega=0)$.
We find that the scattering rate of LSCO is strongly reduced for $T<T_{Neel}$ ($T_{Neel}$ is highlighted by the red vertical arrow),
which is consistent with the kinetic energy reduction observed for LSCO at low temperature (see Fig.~\ref{fig:kinetic_energy}{\bf c}).
The scattering rate of NCCO is much smaller and is only weakly temperature dependent, which
is a consequence of NCCO being less correlated than LSCO.

Hence the location of NCCO and LSCO,
relative to the metal-to-charge-transfer-insulator
boundary \cite{zaanen}, accounts for the observed trends in the
temperature dependence of the optical conductivity. Similar trends of
temperature dependence of the kinetic energy for both electron and hole
doped cuprates were reported in Ref.~\cite{Bontemps}.

\begin{figure}
\begin{center}
\includegraphics[width=\figwidth]{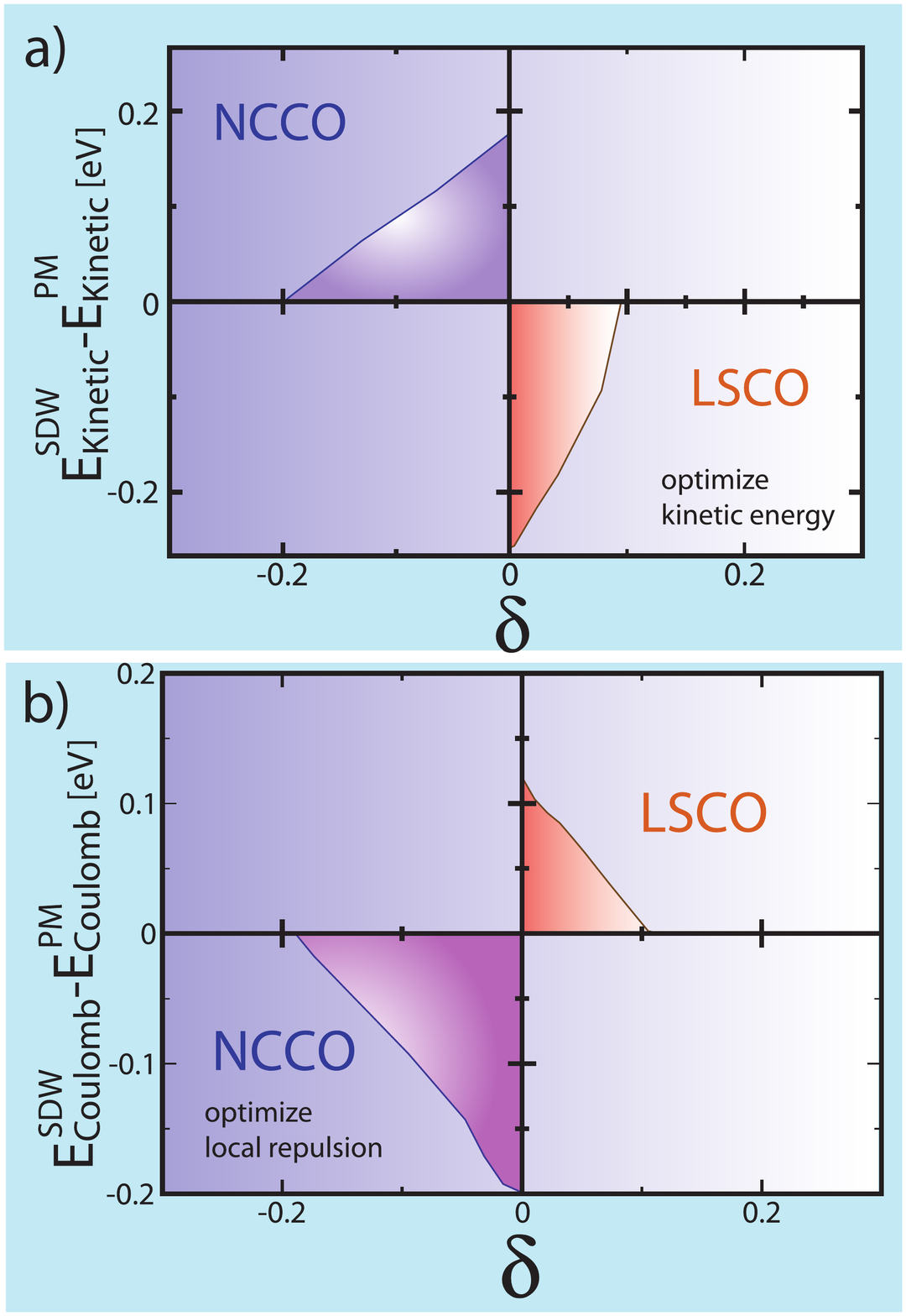}
\caption{
(Colors online) 3-band theoretical energy differences between the antiferromagnetic and the paramagnetic phases.
We show the doping dependence of the a) Kinetic energy $\mathcal{H}_t$ (equation \ref{eq:3band_hub1})
and b) Coulomb energy $\mathcal{H}_U$ (equation \ref{eq:3band_hub2}) of NCCO
and LSCO at fixed temperature T=89K.
There is a kinetic energy optimization when LSCO becomes an antiferromagnet, which is proper to the
Mott insulator, and NCCO is a typical Slater insulator, which optimizes the Coulomb (local onsite repulsion) energy when it becomes
an antiferromagnet, at the expense of a worse Kinetic energy.
This is consistent with the temperature dependence of the theoretical $N_{eff}$ shown in Fig.~\ref{fig:neff_temp} and
with the temperature dependence of the Kinetic and Coulomb energies \ref{fig:kinetic_energy}.
All calculations were obtained by CTQMC.
}
\label{fig:kin_doping}
\end{center}
\end{figure}

Fig.~\ref{fig:kin_doping}\textbf{a} displays the Kinetic energy difference between
the paramagnet and the ordered state, as a function of doping at a fixed temperature T=89K, for NCCO 
and LSCO. The difference of Coulomb energy is shown in Fig.~\ref{fig:kin_doping}\textbf{b}.
We find that in NCCO, the ordered phase is stabilized by optimizing the on-site repulsion (Coulomb energy), at
a cost in the kinetic energy. This is typical for a Slater insulator, and is also captured by a simple Hartree Fock
static mean-field theory. In LSCO the trend is opposite, the ordered phase is stabilized by having a 
lower kinetic energy, at a cost in the Coulomb energy.

Very interestingly, in a one band theory an
increase of the order parameter $\langle S^z \rangle$ leads necessary to a decrease of the Coulomb energy \cite{tremblay_sz} :
\begin{equation}
 \langle \left( S^z \right) ^2 \rangle = \frac{1}{4} \langle \left( n_{\uparrow} - n_{\downarrow} \right) ^2 \rangle  = 
                       \frac{1}{4} \left( n - 2 \langle n_\uparrow n_\downarrow \rangle \right) 
\end{equation}
And therefore :
\begin{equation}
\label{argument_sz}
\langle n_\uparrow n_\downarrow \rangle =  \frac{n}{2} -2\langle \left(S^z\right)^2 \rangle
\end{equation}
In the ordered phase there is hence an increase of $\left(S^z\right)^2$ and 
a decrease of $\langle n_\uparrow n_\downarrow \rangle$ for a fixed density $n$ or doping $\delta$.

In a 3-band theory this is not the case, since the density $n$ in formula (\ref{argument_sz}) is not the total density
but the density of the d orbital $n_d$. The latter quantity is not
fixed at a given doping, and is hence increased when the magnetization is increased.
For the parent compound of LSCO, we found that magnetic correlations lead to an increase of $n_d$ of $1.4\%$, 
and to an increase of the double occupancy $n_{d \up} n_{d \dn}$ of $6\%$.
Hence our results highlight a significant difference between the single band and 3-band theoretical description of LSCO and NCCO.

%%%%%%%%%%%%%%%%%%%%%%%%%%%%%%%%%%%%%%%%%%%%%%%%%%%%%%%%%%%%%%%%
%%%%%%%%%%%%%%%%%%%%%%%%%%%%%%%%%%%%%%%%%%%%%%%%%%%%%%%%%%%%%%%%
%%%%%%%%%%%%%%%%%%%%%%%%%%%%%%%%%%%%%%%%%%%%%%%%%%%%%%%%%%%%%%%%
%%%%%%%%%%%%%%%%%%%%%%%%%%%%%%%%%%%%%%%%%%%%%%%%%%%%%%%%%%%%%%%%
%%%%%%%%%%%%%%%%%%%%%%%%%%%%%%%%%%%%%%%%%%%%%%%%%%%%%%%%%%%%%%%%
%%%%%%%%%%%%%%%%%%%%%%%%%%%%%%%%%%%%%%%%%%%%%%%%%%%%%%%%%%%%%%%%
%%%%%%%%%%%%%%%%%%%%%%%%%%%%%%%%%%%%%%%%%%%%%%%%%%%%%%%%%%%%%%%%
%%%%%%%%%%%%%%%%%%%%%%%%%%%%%%%%%%%%%%%%%%%%%%%%%%%%%%%%%%%%%%%%
%%%%%%%%%%%%%%%%%%%%%%%%%%%%%%%%%%%%%%%%%%%%%%%%%%%%%%%%%%%%%%%%
%%%%%%%%%%%%%%%%%%%%%%%%%%%%%%%%%%%%%%%%%%%%%%%%%%%%%%%%%%%%%%%%
%%%%%%%%%%%%%%%%%%%%%%%%%%%%%%%%%%%%%%%%%%%%%%%%%%%%%%%%%%%%%%%%
%%%%%%%%%%%%%%%%%%%%%%%%%%%%%%%%%%%%%%%%%%%%%%%%%%%%%%%%%%%%%%%%
%%%%%%%%%%%%%%%%%%%%%%%%%%%%%%%%%%%%%%%%%%%%%%%%%%%%%%%%%%%%%%%%
%%%%%%%%%%%%%%%%%%%%%%%%%%%%%%%%%%%%%%%%%%%%%%%%%%%%%%%%%%%%%%%%
%%%%%%%%%%%%%%%%%%%%%%%%%%%%%%%%%%%%%%%%%%%%%%%%%%%%%%%%%%%%%%%%
%%%%%%%%%%%%%%%%%%%%%%%%%%%%%%%%%%%%%%%%%%%%%%%%%%%%%%%%%%%%%%%%
%%%%%%%%%%%%%%%%%%%%%%%%%%%%%%%%%%%%%%%%%%%%%%%%%%%%%%%%%%%%%%%%
%%%%%%%%%%%%%%%%%%%%%%%%%%%%%%%%%%%%%%%%%%%%%%%%%%%%%%%%%%%%%%%%

\section{Strength of correlations in LSCO and NCCO}

We finally extended the 3-band calculations to other values of $\epsilon_d-\epsilon_p$, in order
to study the dependence of our results on the charge transfer energy $\epsilon_d - \epsilon_p$. 
We emphasize that the charge transfer energy plays the role of an effective repulsion U in the one band model language,
and hence controls the strength of the correlations in a 3-band theory. For instance, it was shown in the seminal ZSA paper
\cite{zaanen} that if the Coulomb repulsion of the d orbital is larger than the charger
transfer energy $\epsilon_d-\epsilon_p$,
the size of the gap in the paramagnet is
independent of the Coulomb repulsion, and the strength of correlations is set by the
charge transfer energy.

\begin{figure}
\begin{center}
\includegraphics[width=0.85\columnwidth]{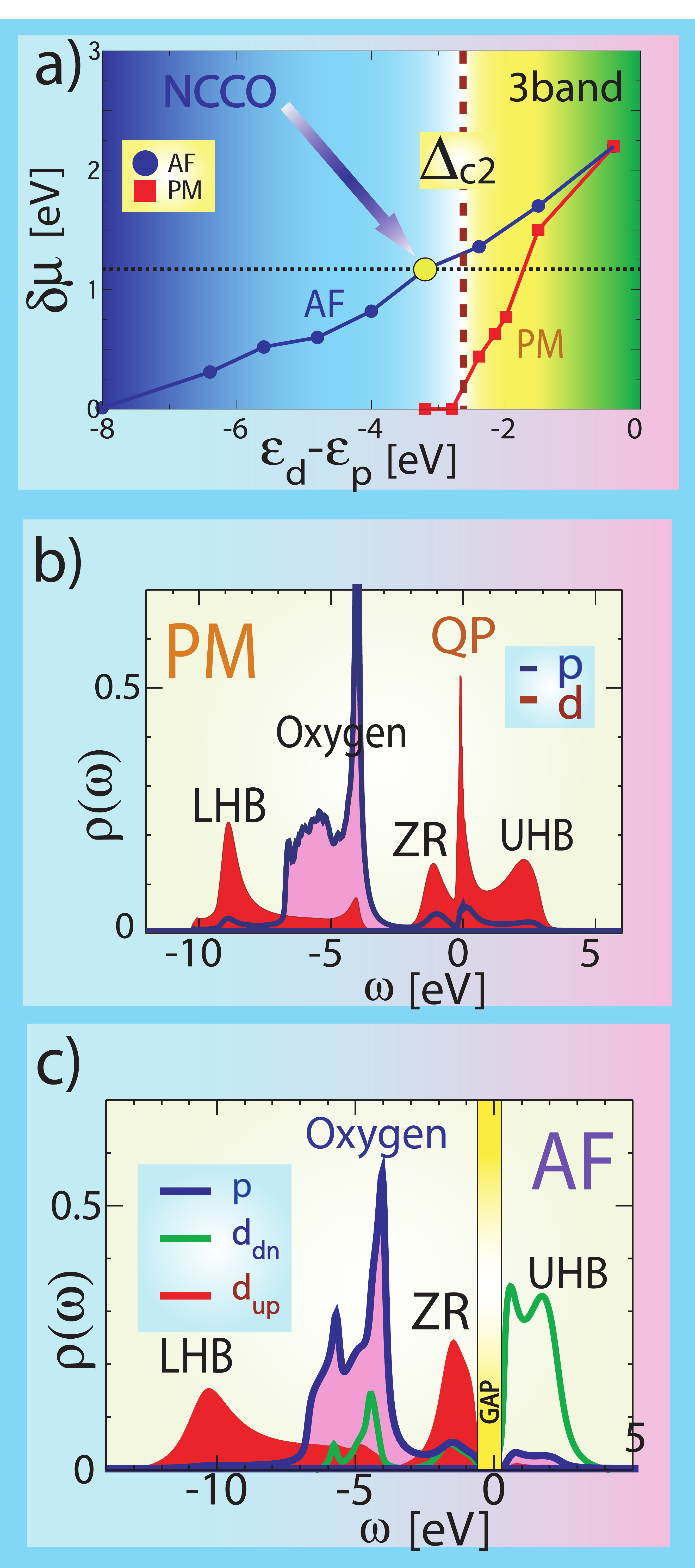}
\caption{
(Colors online)
a) We show the jump in the chemical potential $\delta \mu$
in the ordered state of NCCO (blue circles) for other values of the 
charge transfer energy $\epsilon_d-\epsilon_p$. The results are obtained for the 3-band single site DMFT.
For comparison, we also show the jump in the chemical potential of the paramagnet (red squares).
There is a quantum critical point $\Delta_{c2}$ for the paramagnetic state, with respect to
the charge transfer energy, corresponding to the metal to charge transfer insulator transition. 
The physical value obtained by LDA+DMFT for  $\epsilon_d-\epsilon_p$ places NCCO below the boundary. 
b) Density of states of the paramagnetic state of NCCO. The quasi-particle peak close to the Fermi
energy (QP) is a signature that NCCO is paramagnetic metal.
c) Density of states of the ordered state of NCCO.
}
\label{phasediag_ncco}
\end{center}
\end{figure}

\begin{figure}
\begin{center}
\includegraphics[width=0.85\columnwidth]{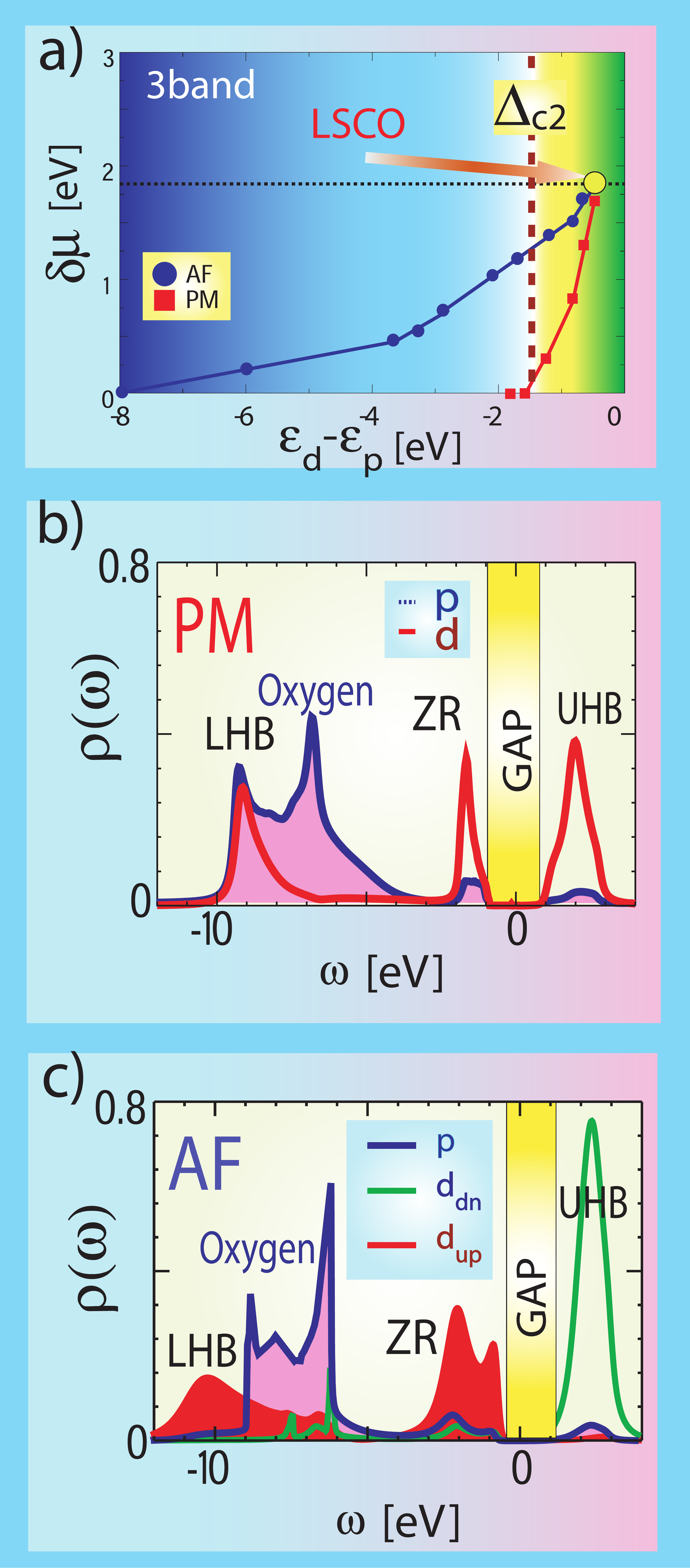}
\caption{
(Colors online)
a) We show the jump in the chemical potential $\delta \mu$
in the ordered state of LSCO (blue circles) for other values of
charge transfer energy $\epsilon_d-\epsilon_p$. The results are obtained for the 3-band single site DMFT.
For comparison, we also show the jump in the chemical potential of the paramagnet (red squares).
There is a quantum critical point $\Delta_{c2}$ for the paramagnetic state, with respect to
the charge transfer energy, corresponding to the metal to charge transfer
insulator transition.  The physical value obtained by LDA+DMFT for $\epsilon_d-\epsilon_p$ place LSCO above the boundary.
b) Density of states of the paramagnetic state of LSCO, which shows the presence of a gap. This is
a signature that LSCO is a Mott insulator
c) Density of states of the ordered state of LSCO. The gap in the density of states is of similar size
    for the paramagnetic insulator and the ordered state as shown in details in Fig.~\ref{fig:lsco_pm_sdw}.
}
\label{phasediag}
\end{center}
\end{figure}

In order to study the strength of the correlations, we
computed the jump in the chemical potential $\delta \mu$ 
for infinitesimal doping of the parent compound.
This quantity gives an estimation for the gap in the spectral functions 
of the parent compound, which is around 1.2 eV and 1.8 eV in NCCO and LSCO
respectively.
We present the data for both the ordered state and the paramagnetic state of LSCO and NCCO
in  Figs.~\ref{phasediag_ncco}{\bf a}) and ~\ref{phasediag}{\bf a}), respectively.
The jump of chemical potential in the paramagnet gives  
an estimation of the critical charge transfer energy $\Delta_{c2}$, 
which is the maximum charge transfer energy that allows the metallic solution.
However at $\Delta_{c2}$ the ordered solution has a substantial gap, which
is closely related to $\delta \mu$ in the ordered state. 
We find that $\delta \mu$ in the ordered state of NCCO is around 1.2 eV and 1.8 eV
in LSCO.
Hence, we conclude that NCCO (\ref{phasediag_ncco}.a) is slightly below the charge-transfer-insulator-to-metal transition $\Delta_{c2}$.
Indeed, below $\Delta_{c2}$, the magnetic long-range correlations are necessary to open a gap, and slightly above $\Delta_{c2}$
the paramagnetic gap is much smaller than the slater gap induced by the nesting, 
as shown in Fig.~\ref{phasediag_ncco}{\bf a}.

The two solutions of the DMFT equations for NCCO are 
shown in panels \ref{phasediag_ncco}\textbf{b,c}.
The first one is non-magnetic and metallic, and describes a material in the absence of long range
order. The second is insulating and antiferromagnetically
ordered, with a charge transfer gap of $1.2$eV. Since the non-magnetic solution is metallic, the magnetic long range
order is responsible for the insulating nature of NCCO (Slater insulator).

We find that LSCO (Fig.~\ref{phasediag}{\bf a}) is above $\Delta_{c2}$, as 
reported in a recent study \cite{our_previous_paper_lsco}.
Indeed, the parent compound of LSCO is only weakly affected by the presence of magnetic order,
the size of the gap is only slightly increased when magnetic order is present.
The DMFT equations for LSCO have two solutions, shown in panels \ref{phasediag}\textbf{b,c}.
The first one is paramagnetic and the second is antiferromagnetically
ordered, with a charge transfer gap of $1.8$eV. Since the paramagnetic solution is insulating, the magnetic long range
order is not responsible for the insulating nature of LSCO (Mott insulator).

\begin{figure}
\begin{center}
\includegraphics[width=\miniwidth]{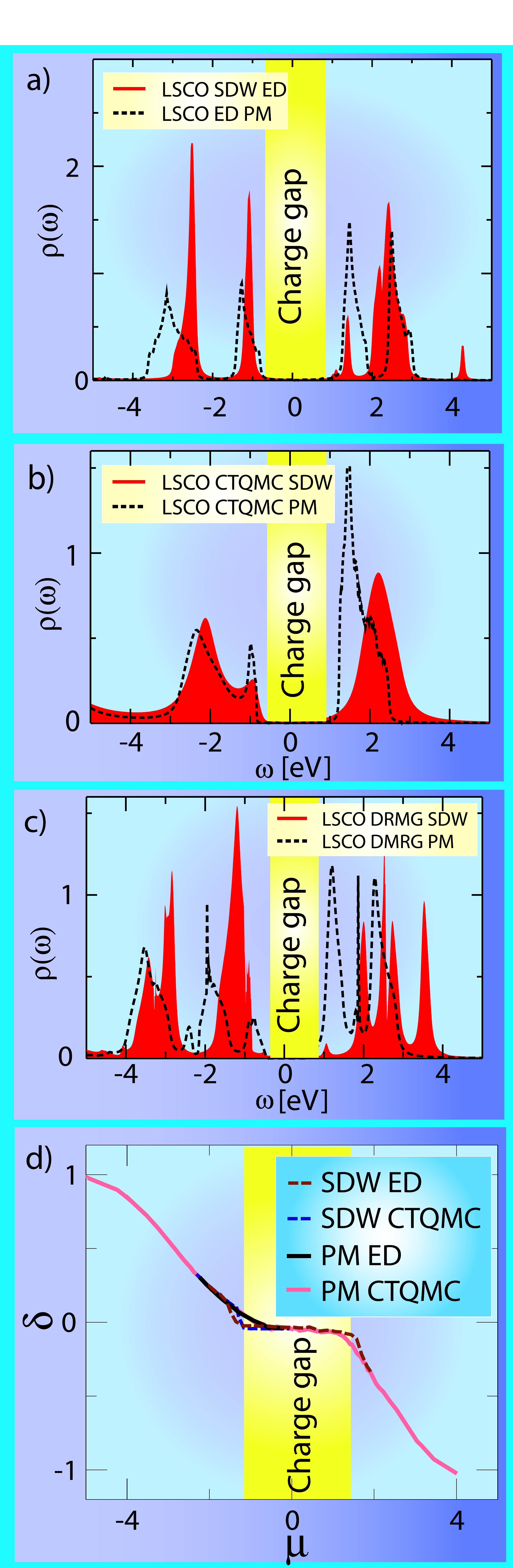}
\caption{
  (Colors online)
   Density of states of the 3-band description of the parent compound of LSCO in the paramagnetic (dashed line) and ordered state (red area) obtained
   with different solvers :  a) exact diagonalization (ED), b) CTQMC and c) density matrix renormalization group (DMRG).
   All data show that LSCO is a paramagnetic insulator, and that the size of the gap obtained by the density of states is similar for both
   the paramagnet and the ordered state (within $\approx 10\%$). d) We show the variation of the doping $\delta$
   with respect to the chemical potential $\mu$ for both the ordered state (SDW) and the paramagnetic state (PM) obtained
   by exact diagonalization (ED) and CTQMC. There is a jump in the chemical potential $\delta \mu$ (plateau at $\delta=0$)
   of similar size for all data. This shows that the magnetic correlations
   do not strongly affect the insulating properties of LSCO.
}
\label{fig:lsco_pm_sdw}
\end{center}
\end{figure}

In Fig.~\ref{fig:lsco_pm_sdw}{\bf a-c}, we show the density of states of the 3-band description of the parent compound of LSCO using various
numerical tools. In panel \textbf{a} we show the density of states obtained by ED for the paramagnet and the ordered states.
In panel \textbf{b} we show the density of states obtained by CTQMC and in panel \textbf{c} the density
of states obtained by a recent DMRG solver \cite{amaricci_dmrg}. 
We conclude that the gap in the ordered state of LSCO is of similar
size than the gap obtained in the paramagnet, independently of the numerical solver used to solve the DMFT equations.
In Fig.~\ref{fig:lsco_pm_sdw}{\bf d} we show the doping as a function of the chemical potential, and it displays a plateau
related to the charge gap at integer filling. 
The agreement between the different solvers gives us confidence in these results.

The asymmetry between both NCCO and LSCO, being 
below and above the charge-transfer-insulator-to-metal transition $\Delta_{c2}$, 
is a simple explanation for the striking asymmetry in their phase diagram. 
For LSCO, the magnetic correlations are destroyed rapidly upon doping, while in the NCCO they
survive up to high doping, as shown in Fig.~\ref{fig:magnetism}{\bf a}.

In this section, we emphasized that the physical origin of the asymmetry between LSCO and
NCCO lies not only in the different values of the oxygen-oxygen
overlap, which controls the curvature of the Fermi surface, an effect
that is captured in model Hamiltonian studies, but also in the different
values of the charge transfer gap in the two structures. The latter
has an electrostatic origin, the electron doped material lacks the
negatively charged apical oxygen, which increase the electrostatic
potential at the copper site.

\begin{figure}
\begin{center}
\includegraphics[width=\figwidth]{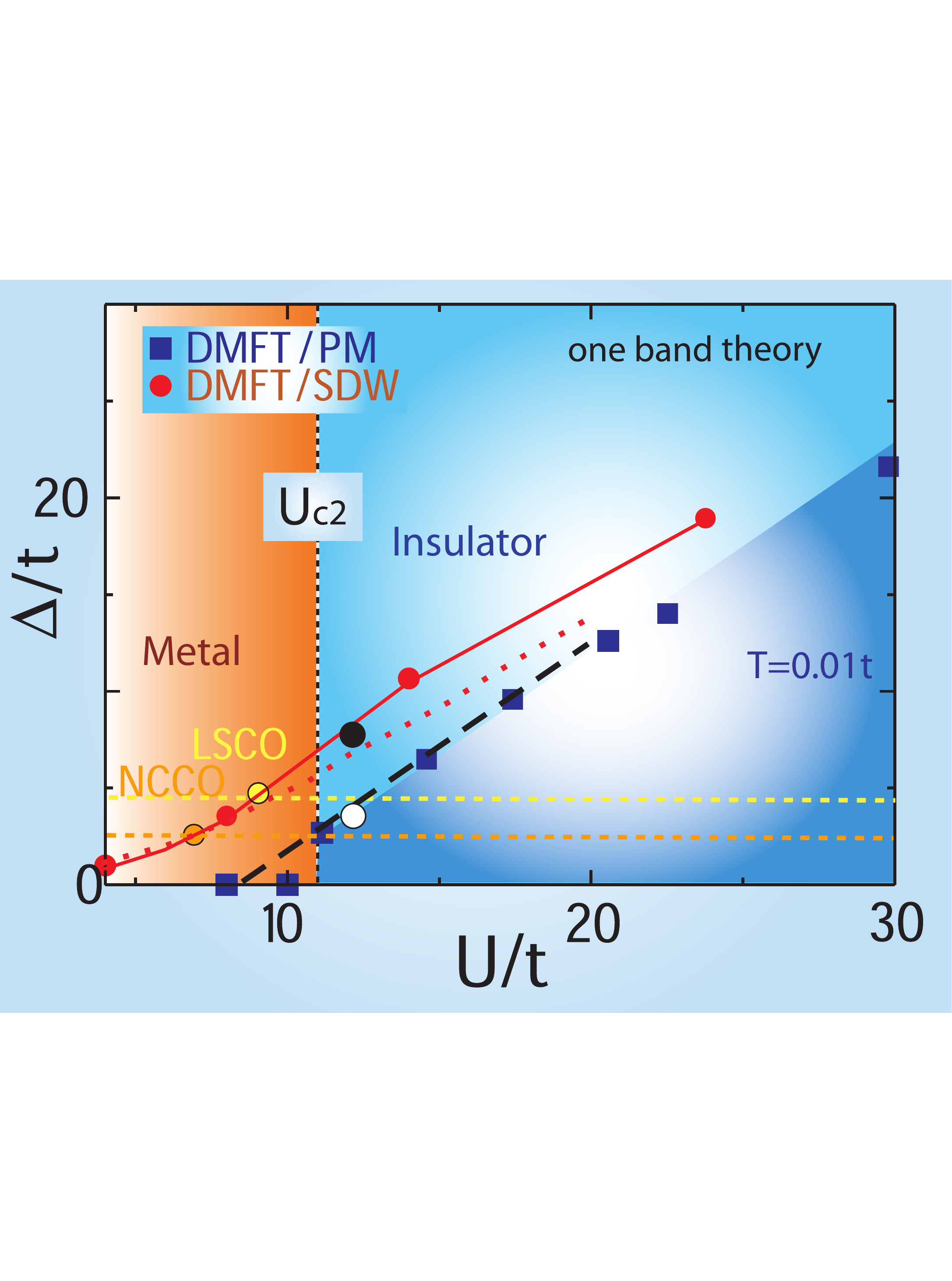}
\caption{
(Colors online)
We show calculations done on the one band Hubbard model for various Coulomb repulsion
$U$ and transfer integral $t$ ratios. In the one band theory there is also a quantum critical point $U_{c2}$
that is the minimal repulsion driving an paramagnet to an insulator. Locating the compounds LSCO and NCCO
by fitting the gap $\Delta$ ($\Delta=1.2$eV and $t=0.42$eV \cite{oneband} for NCCO and $\Delta=1.8$eV and $t=0.43$eV for LSCO),
we place both LSCO and NCCO below $U_{c2}$ in the one band picture. It is worth noting that for
large enough $U/t$ the size of the paramagnetic gap is close to the gap in the ordered state.
Results were obtained by using single site DMFT with exact diagonalization.
For comparison we show theoretical calculations of Ref. \cite{rosenberg_1band}, done for
the paramagnet (long dashed line) and the ordered state (short dotted line), and the theoretical
calculations of Ref. \cite{millis_compound_at_uc2} in the paramagnetic state (white circle) and in 
the ordered state (black circle).
}
\label{phasediag_oneband}
\end{center}
\end{figure}

For comparison, we now also discuss results of the one band Hubbard model.
In Fig.~\ref{phasediag_oneband} we show the gap in the density of states for the
paramagnetic insulator and for the antiferromagnetic insulator.
In the one band model exists a critical point $U_{c2}$
which separates the paramagnetic metal at small $U/t$ 
from the paramagnetic insulator at large $U/t$ (squares). 
The magnetic solution is always insulating (circles). Using typical values
for the hopping parameter $t$ \cite{oneband}($t=0.42$eV for NCCO and $t=0.43$eV for LSCO), 
and typical values for the gap $\Delta$ in the ordered
state for LSCO and NCCO ($\Delta=1.2$eV for NCCO and $\Delta=1.8$eV for LSCO) 
we can locate the compounds in the one-band model phase diagram. 
Both LSCO and NCCO are below $U_{c2}$ in this picture, in agreement with Ref \cite{luca_nature}.
There is therefore a strong difference in the physical conclusions 
obtained by the one-band calculations and the LDA+DMFT. 

%%%%%%%%%%%%%%%%%%%%%%%%%%%%%%%%%%%%%%%%%%%%%%%%%%%%%%%%%%%%%%%%
%%%%%%%%%%%%%%%%%%%%%%%%%%%%%%%%%%%%%%%%%%%%%%%%%%%%%%%%%%%%%%%%
%%%%%%%%%%%%%%%%%%%%%%%%%%%%%%%%%%%%%%%%%%%%%%%%%%%%%%%%%%%%%%%%
%%%%%%%%%%%%%%%%%%%%%%%%%%%%%%%%%%%%%%%%%%%%%%%%%%%%%%%%%%%%%%%%
%%%%%%%%%%%%%%%%%%%%%%%%%%%%%%%%%%%%%%%%%%%%%%%%%%%%%%%%%%%%%%%%
%%%%%%%%%%%%%%%%%%%%%%%%%%%%%%%%%%%%%%%%%%%%%%%%%%%%%%%%%%%%%%%%
%%%%%%%%%%%%%%%%%%%%%%%%%%%%%%%%%%%%%%%%%%%%%%%%%%%%%%%%%%%%%%%%
%%%%%%%%%%%%%%%%%%%%%%%%%%%%%%%%%%%%%%%%%%%%%%%%%%%%%%%%%%%%%%%%
%%%%%%%%%%%%%%%%%%%%%%%%%%%%%%%%%%%%%%%%%%%%%%%%%%%%%%%%%%%%%%%%
%%%%%%%%%%%%%%%%%%%%%%%%%%%%%%%%%%%%%%%%%%%%%%%%%%%%%%%%%%%%%%%%
%%%%%%%%%%%%%%%%%%%%%%%%%%%%%%%%%%%%%%%%%%%%%%%%%%%%%%%%%%%%%%%%
%%%%%%%%%%%%%%%%%%%%%%%%%%%%%%%%%%%%%%%%%%%%%%%%%%%%%%%%%%%%%%%%
%%%%%%%%%%%%%%%%%%%%%%%%%%%%%%%%%%%%%%%%%%%%%%%%%%%%%%%%%%%%%%%%
%%%%%%%%%%%%%%%%%%%%%%%%%%%%%%%%%%%%%%%%%%%%%%%%%%%%%%%%%%%%%%%%
%%%%%%%%%%%%%%%%%%%%%%%%%%%%%%%%%%%%%%%%%%%%%%%%%%%%%%%%%%%%%%%%
%%%%%%%%%%%%%%%%%%%%%%%%%%%%%%%%%%%%%%%%%%%%%%%%%%%%%%%%%%%%%%%%
%%%%%%%%%%%%%%%%%%%%%%%%%%%%%%%%%%%%%%%%%%%%%%%%%%%%%%%%%%%%%%%%
%%%%%%%%%%%%%%%%%%%%%%%%%%%%%%%%%%%%%%%%%%%%%%%%%%%%%%%%%%%%%%%%

\section{\bf Conclusion}

In conclusion, we carried out a comparative study of NCCO and
LSCO using a realistic LDA+DMFT approach. The LDA+DMFT studies 
achieved overall good agreement with optical
conductivity and ARPES studies in a broad range of dopings and a
wide range of energy scales up to energies of the order of the
charge transfer energy.
 
In particular, for NCCO we found that static mean-field 
theory is not sufficient to describe many qualitative 
features due to the presence of
multiple peaks in the electronic spectra. The description of
these features require more sophisticated methods and studies
along those lines were carried out for NCCO in Refs
\cite{bansil1, bansil2}. We demonstrated that LDA+DMFT
successfully describes these effects in NCCO.

While single site DMFT is already a good methodology 
to describe the phase diagram of NCCO, cluster corrections are important in
the underdoped region of LSCO. This indicates the importance of
singlet formation or the possible importance of other ordered
states in this region. However, our LDA+DMFT gave a remarkable agreement with experiments 
for the optical conductivity, for doping smaller than $20\%$ and energy scale $\omega<2$eV,
when the \dz and \pz are considered.
For larger doping and higher frequencies, additional bands must should considered for 
a proper description of the photoemission spectra.

Moreover, in agreement with Ref \cite{more_ref21} we found that even after
inclusion of the apical oxygens, single site DMFT does not
capture the saturation of the occupancy observed around doping 0.2 in
the x-ray absorption spectroscopy (XAS) 
experiments of Ref. \cite{xray_absorption}. We notice however
that LDA+DMFT does capture the evolution of the {\it  ratio
of the occupancies} of apical and planar oxygens, including a
rapid increase of the occupancy of $p_z$ around doping 0.2. 
We note that modeling XAS more accurately may require:
i) Taking into account additional LDA bands and more orbitals,
ii) To include the doping dependence of the apical oxygen position pointed
out in Ref \cite{more_ref16}, 
iii) Non-local correlations beyond DMFT might be relevant as suggested in 
     Refs \onlinecite{more_ref19,more_ref20,more_ref21,more_ref22},
iv) To include the core hole potential, as it was done for the 
    core level photoemission in Ref \cite{cornaglia}, might also be necessary.

We achieved a successful description of many normal state physical properties
(optical conductivity, ARPES, stability of magnetism) of these two archetypal cuprates. 
The overall quantitative agreement between LDA+DMFT and experiments,
for both LSCO and NCCO gives us
increased confidence in LDA+DMFT as an approach to strongly correlated materials.

We emphasize that it did not require the introduction of ad hoc parameters, 
such as a doping dependent interaction strength. 
The sensitivity of the results on the choice
of the charge transfer energy, which essentially determines
the strength of the correlations, was explored.
We emphasize that the charge transfer energy for NCCO and LSCO is obtained from
ab initio calculations, and for the values obtained by first principle
calculations we obtained good agreement with experiments.

It is remarkable to obtain good quantitative agreement
between theory and experiment, on energy scales of the order the
charge transfer energy. For example, our methodology gives a 
good agreement with experiments for the integrated
spectral weight of the optical conductivity with a cutoff above
the charge transfer gap in LSCO. 
For this quantitative agreement it is essential to use a
multiband model including the apical oxygen orbitals. The fact
that these are absent in NCCO accounts for the good agreement
with the magnitude of the optical conductivity obtained for this
material in our earlier publication \cite{our_nature_paper}.

Moreover, our methodology applied to two different materials captures not only
their similarities, as for example the doping dependence of the 
integrated spectral weight,
but  also  their differences, as the doping dependence of the magnetic moment.
These differences stems from their different electronic structure and is well
captured by LDA+DMFT.
Differences between the electron and hole doped cuprates have
been noticed by many authors. They were interpreted as arising
from both the bare hopping integrals in the one band model Hamiltonian 
and the Coulomb interaction U in the Hubbard model.

However, LDA+DMFT and multi-band theories are also able to  
capture a more fundamental difference resulting from the
different strength of the correlations in these two materials, 
which is driven by the charge transfer energy.
Using our methodology, we determined the relative
strength of the  correlations of LSCO and NCCO. 
LSCO was found to be on the insulating side of the
ZSA phase boundary confirming the preliminary conclusions of reference
\cite{our_previous_paper_lsco}, but in disagreement with Ref
\cite{luca_nature} which classifies all the parent compound of
the copper oxides as Slater insulators. The electron doped
compound NCCO was found to be on the metallic side of the
ZSA phase boundary \cite{our_nature_paper} in
agreement with  ref \cite{luca_nature}, but in disagreement with
Ref \cite{gavrichkov}.
Indeed, the lack of apical oxygens in NCCO reduces the charge transfer
energy relative to LSCO, placing these two materials on two
different sides of the ZSA phase boundary.

This work is complementary to our earlier work \cite{our_nature_paper}, placing the
copper oxides in a region of intermediate correlation strength.
For materials in this region of
parameters, we found that the location of the material relative
to the ZSA boundary has important physical consequences. For
example the evolutions of the optical properties with
doping and temperature in NCCO and LSCO are different,
in agreement with the earlier work of Ref~\cite{nicole_bontemps}.
Our results allowed to shed some light on the differences
observed in LSCO and NCCO, and these differences are attributed 
to the location of LSCO and NCCO on the two different sides of the ZSA
boundary.

This has direct consequences on physical observables at finite doping.
For instance, antiferromagnetism in the Slater limit is accompanied by an
increase in the expectation value of the kinetic energy (with a
concomitant reduction in the double occupation) while in the Mott
limit the insulating state has an optimization of the kinetic energy which is a
manifestation of the increase of the expectation
value of the superexchange interaction. These is
a transparent interpretation of the different trends in the
evolution of the optical spectroscopy in NCCO and LSCO materials.

Our method captures both the similarities and the many essential differences
between the two compounds, which can be traced to their structure
and atomic constituents, and in particular to the absence of
apical oxygens in the T' structure of NCCO. 
However, there are still avenues to improve the single site DMFT
description of copper oxide  materials, by including the
effects of the nearest neighbor Coulomb interaction between
copper and oxygen and among the oxygens beyond the Hartree
approximation, and inclusion of frequency dependent screening.
Many local or frequency integrated quantities are already well
described by single site DMFT, but in the region of hole doping
where the discrepancy between single site DMFT and cluster DMFT is
noticeable, the latter should be used to refine the description
of low energy physical properties.

An important direction, to be pursued is the
study of other ordered states, that exist as stable or metastable
solutions of the realistic single site or cluster LDA+DMFT
equations. Exploration of this landscape of DMFT solutions, is a
challenging project, and is worthwhile to pursue in conjunction with studies of the
superconducting phase. These problems are currently under investigation.

%%%%%%%%%%%%%%%%%%%%%%%%%%%%%%%%%%%%%%%%%%%%%%%%%%%%%%%%%%%%%%%%
%%%%%%%%%%%%%%%%%%%%%%%%%%%%%%%%%%%%%%%%%%%%%%%%%%%%%%%%%%%%%%%%
%%%%%%%%%%%%%%%%%%%%%%%%%%%%%%%%%%%%%%%%%%%%%%%%%%%%%%%%%%%%%%%%
%%%%%%%%%%%%%%%%%%%%%%%%%%%%%%%%%%%%%%%%%%%%%%%%%%%%%%%%%%%%%%%%
%%%%%%%%%%%%%%%%%%%%%%%%%%%%%%%%%%%%%%%%%%%%%%%%%%%%%%%%%%%%%%%%
%%%%%%%%%%%%%%%%%%%%%%%%%%%%%%%%%%%%%%%%%%%%%%%%%%%%%%%%%%%%%%%%

\section{Acknowledgement}

We thank A.M. Tremblay, D. Basov, D. G. Hawthorn, G. A. Sawatzky 
for discussions and sharing their insights and experimental data. Numerous
discussions with A. Georges,  A. Amaricci, J. C. Domenge and A.
Millis are gratefully acknowledged. Adriano Amaricci shared his
density matrix renormalization group code and Jean-Christophe
Dommenge shared his exact diagonalization code. K.H was
supported by Grant NSF NFS DMR-0746395 and Alfred P. Sloan
fellowship. G.K. was supported by NSF DMR-0906943, and C.W. was
supported by the Swiss National Foundation for Science (SNFS).

%%%%%%%%%%%%%%%%%%%%%%%%%%%%%%%%%%%%%%%%%%%%%%%%%%%%%%%%%%%%%%%%
%%%%%%%%%%%%%%%%%%%%%%%%%%%%%%%%%%%%%%%%%%%%%%%%%%%%%%%%%%%%%%%%
%%%%%%%%%%%%%%%%%%%%%%%%%%%%%%%%%%%%%%%%%%%%%%%%%%%%%%%%%%%%%%%%
%%%%%%%%%%%%%%%%%%%%%%%%%%%%%%%%%%%%%%%%%%%%%%%%%%%%%%%%%%%%%%%%
%%%%%%%%%%%%%%%%%%%%%%%%%%%%%%%%%%%%%%%%%%%%%%%%%%%%%%%%%%%%%%%%
%%%%%%%%%%%%%%%%%%%%%%%%%%%%%%%%%%%%%%%%%%%%%%%%%%%%%%%%%%%%%%%%
%%%%%%%%%%%%%%%%%%%%%%%%%%%%%%%%%%%%%%%%%%%%%%%%%%%%%%%%%%%%%%%%
%%%%%%%%%%%%%%%%%%%%%%%%%%%%%%%%%%%%%%%%%%%%%%%%%%%%%%%%%%%%%%%%
%%%%%%%%%%%%%%%%%%%%%%%%%%%%%%%%%%%%%%%%%%%%%%%%%%%%%%%%%%%%%%%%
%%%%%%%%%%%%%%%%%%%%%%%%%%%%%%%%%%%%%%%%%%%%%%%%%%%%%%%%%%%%%%%%
%%%%%%%%%%%%%%%%%%%%%%%%%%%%%%%%%%%%%%%%%%%%%%%%%%%%%%%%%%%%%%%%
%%%%%%%%%%%%%%%%%%%%%%%%%%%%%%%%%%%%%%%%%%%%%%%%%%%%%%%%%%%%%%%%
%%%%%%%%%%%%%%%%%%%%%%%%%%%%%%%%%%%%%%%%%%%%%%%%%%%%%%%%%%%%%%%%
%%%%%%%%%%%%%%%%%%%%%%%%%%%%%%%%%%%%%%%%%%%%%%%%%%%%%%%%%%%%%%%%
%%%%%%%%%%%%%%%%%%%%%%%%%%%%%%%%%%%%%%%%%%%%%%%%%%%%%%%%%%%%%%%%
%%%%%%%%%%%%%%%%%%%%%%%%%%%%%%%%%%%%%%%%%%%%%%%%%%%%%%%%%%%%%%%%
%%%%%%%%%%%%%%%%%%%%%%%%%%%%%%%%%%%%%%%%%%%%%%%%%%%%%%%%%%%%%%%%
%%%%%%%%%%%%%%%%%%%%%%%%%%%%%%%%%%%%%%%%%%%%%%%%%%%%%%%%%%%%%%%%
%%%%%%%%%%%%%%%%%%%%%%%%%%%%%%%%%%%%%%%%%%%%%%%%%%%%%%%%%%%%%%%%
%%%%%%%%%%%%%%%%%%%%%%%%%%%%%%%%%%%%%%%%%%%%%%%%%%%%%%%%%%%%%%%%
%%%%%%%%%%%%%%%%%%%%%%%%%%%%%%%%%%%%%%%%%%%%%%%%%%%%%%%%%%%%%%%%
%%%%%%%%%%%%%%%%%%%%%%%%%%%%%%%%%%%%%%%%%%%%%%%%%%%%%%%%%%%%%%%%

\section{Appendix A}
\label{bz_folding}

In this appendix, we discuss how the spectral weight obtained in the ordered state, $A^\alpha(\bold{K},\omega)$ 
is mapped to the unfolded Brillouin zone for comparison with experiments.

For the calculations done in the ordered state, $(A^\alpha(\bold{K},\omega))^{ij}$ is obtained in the folded Brillouin Zone, and
therefore is now labeled by two additional indices $i,j$, that are
running indices over the paramagnetic unitcell block of the extended unitcell.

The corresponding spectral weight can be obtained in the unfolded Brillouin zone by the relation:
\begin{equation}
A(\bold{k},\omega)^\alpha = \sum\limits_{i,j=1}^{N_{cell}}{ e^{i\bold{k}\left(\bold{R}_i-\bold{R}_j \right)} \left(\bold{A}^\alpha(\bold{K},\omega)\right)^{ij} }
\end{equation}
Where $N_{cell}$ is the number of paramagnetic unitcell contained in the extended ordered state unitcell, and $\bold{R}_{i,j}$ is the position
of the block within the extended unitcell.

The relation between $k$ (unfolded Brillouin zone, paramagnetic state) and $K$ (folded Brillouin zone, ordered state) is given by: 
\begin{equation*}
 \bold{k}=\left( K_x A_1 + K_y B_1 \right) \bold{g}_1 + \left( K_x A_2 + K_y B_2 \right)  \bold{g}_2 
\end{equation*}
With the following definitions. $\bold{K}=K_x \bold{G}_1+K_y \bold{G}_2$, where $G_{1,2}$ are the reciprocal basis vector
of the ordered state, $\bold{a}= A_1 \bold{E}_1 + B_1 \bold{E}_2$ and $\bold{b}= A_2 \bold{E}_1 + B_2 \bold{E}_2$, 
where $\bold{a,b}$ are the direct space basis vector of the paramagnetic unitcell, and $\bold{E}_{1,2}$ are the direct space
basis vector of the extended state unitcell (ordered state).

\section{Appendix B}
\label{hopping}

In this section we show the realistic set of tight binding parameters obtained by downfolding
the LDA band structure (see Table \ref{table:lda_param_lsco} and \ref{table:lda_param_ncco}).

\begin{table}     
 \caption{3-band model Hamiltonian parameters obtained by LDA downfolding for LSCO.
 The vector connecting the     
 two different unit-cells is shown, and the two orbitals     
 connected by the hopping as well.     
 }     
 \begin{tabular}{|c|c|c||c|c|c|}     
   \hline     
     orbitals & vector &  amplitude & orbitals & vector & amplitude  \\     
   \hline     
     $d_x$-$d_x$ &           (-2,0 ) &  -0.01 &      $d_x$-$p_y$ &           (-2,0 ) &  
 -0.01 \\
     $p_y$-$p_y$ &           (-2,0 ) &  0.010 &      $d_x$-$d_x$ &           (-1,-1) &  
 0.030 \\
     $d_x$-$p_x$ &           (-1,-1) &  -0.03 &      $d_x$-$p_y$ &           (-1,-1) &  
 0.030 \\
     $p_x$-$p_x$ &           (-1,-1) &  0.030 &      $p_y$-$p_y$ &           (-1,-1) &  
 0.030 \\
     $d_x$-$d_x$ &           (-1,0 ) &  0.030 &      $d_x$-$p_x$ &           (-1,0 ) &  
 0.030 \\
     $d_x$-$p_y$ &           (-1,0 ) &  -1.40 &      $p_x$-$d_x$ &           (-1,0 ) &  
 0.030 \\
     $p_x$-$p_x$ &           (-1,0 ) &  -0.03 &      $p_x$-$p_y$ &           (-1,0 ) &  
 0.660 \\
     $p_y$-$d_x$ &           (-1,0 ) &  0.010 &      $p_y$-$p_y$ &           (-1,0 ) &  
 0.030 \\
     $d_x$-$d_x$ &           (-1,1 ) &  0.030 &      $d_x$-$p_y$ &           (-1,1 ) &  
 0.030 \\
     $p_x$-$d_x$ &           (-1,1 ) &  -0.03 &      $p_x$-$p_x$ &           (-1,1 ) &  
 0.030 \\
     $p_x$-$p_y$ &           (-1,1 ) &  -0.66 &      $p_y$-$p_y$ &           (-1,1 ) &  
 0.030 \\
     $d_x$-$d_x$ &           (0 ,-2) &  -0.01 &      $d_x$-$p_x$ &           (0 ,-2) &  
 0.010 \\
     $p_x$-$p_x$ &           (0 ,-2) &  0.010 &      $p_y$-$p_y$ &           (0 ,-2) &  
 -0.01 \\
     $d_x$-$d_x$ &           (0 ,-1) &  0.030 &      $d_x$-$p_x$ &           (0 ,-1) &  
 1.400 \\
     $d_x$-$p_y$ &           (0 ,-1) &  -0.03 &      $p_x$-$d_x$ &           (0 ,-1) &  
 -0.01 \\
     $p_x$-$p_x$ &           (0 ,-1) &  0.030 &      $p_y$-$d_x$ &           (0 ,-1) &  
 -0.03 \\
     $p_y$-$p_x$ &           (0 ,-1) &  0.660 &      $p_y$-$p_y$ &           (0 ,-1) &  
 -0.03 \\
     $d_x$-$d_x$ &           (0 ,0 ) &  10.87 &      $d_x$-$p_x$ &           (0 ,0 ) &  
 -1.40 \\
     $d_x$-$p_y$ &           (0 ,0 ) &  1.400 &      $p_x$-$d_x$ &           (0 ,0 ) &  
 -1.40 \\
     $p_x$-$p_x$ &           (0 ,0 ) &  8.110 &      $p_x$-$p_y$ &           (0 ,0 ) &  
 -0.66 \\
     $p_y$-$d_x$ &           (0 ,0 ) &  1.400 &      $p_y$-$p_x$ &           (0 ,0 ) &  
 -0.66 \\
     $p_y$-$p_y$ &           (0 ,0 ) &  8.110 &      $d_x$-$d_x$ &           (0 ,1 ) &  
 0.030 \\
     $d_x$-$p_x$ &           (0 ,1 ) &  -0.01 &      $d_x$-$p_y$ &           (0 ,1 ) &  
 -0.03 \\
     $p_x$-$d_x$ &           (0 ,1 ) &  1.400 &      $p_x$-$p_x$ &           (0 ,1 ) &  
 0.030 \\
     $p_x$-$p_y$ &           (0 ,1 ) &  0.660 &      $p_y$-$d_x$ &           (0 ,1 ) &  
 -0.03 \\
     $p_y$-$p_y$ &           (0 ,1 ) &  -0.03 &      $d_x$-$d_x$ &           (0 ,2 ) &  
 -0.01 \\
     $p_x$-$d_x$ &           (0 ,2 ) &  0.010 &      $p_x$-$p_x$ &           (0 ,2 ) &  
 0.010 \\
     $p_y$-$p_y$ &           (0 ,2 ) &  -0.01 &      $d_x$-$d_x$ &           (1 ,-1) &  
 0.030 \\
     $d_x$-$p_x$ &           (1 ,-1) &  -0.03 &      $p_x$-$p_x$ &           (1 ,-1) &  
 0.030 \\
     $p_y$-$d_x$ &           (1 ,-1) &  0.030 &      $p_y$-$p_x$ &           (1 ,-1) &  
 -0.66 \\
     $p_y$-$p_y$ &           (1 ,-1) &  0.030 &      $d_x$-$d_x$ &           (1 ,0 ) &  
 0.030 \\
     $d_x$-$p_x$ &           (1 ,0 ) &  0.030 &      $d_x$-$p_y$ &           (1 ,0 ) &  
 0.010 \\
     $p_x$-$d_x$ &           (1 ,0 ) &  0.030 &      $p_x$-$p_x$ &           (1 ,0 ) &  
 -0.03 \\
     $p_y$-$d_x$ &           (1 ,0 ) &  -1.40 &      $p_y$-$p_x$ &           (1 ,0 ) &  
 0.660 \\
     $p_y$-$p_y$ &           (1 ,0 ) &  0.030 &      $d_x$-$d_x$ &           (1 ,1 ) &  
 0.030 \\
     $p_x$-$d_x$ &           (1 ,1 ) &  -0.03 &      $p_x$-$p_x$ &           (1 ,1 ) &  
 0.030 \\
     $p_y$-$d_x$ &           (1 ,1 ) &  0.030 &      $p_y$-$p_y$ &           (1 ,1 ) &  
 0.030 \\
     $d_x$-$d_x$ &           (2 ,0 ) &  -0.01 &      $p_y$-$d_x$ &           (2 ,0 ) &  
 -0.01 \\
     $p_y$-$p_y$ &           (2 ,0 ) &  0.010 &      $p_y$-$p_y$ &           (2 ,2 ) &  
 0.000 \\
   \hline     
   \hline     
 \end{tabular}     
 \label{table:lda_param_lsco}     
 \end{table}

 \begin{table}     
 \caption{3-band model Hamiltonian parameters obtained by LDA downfolding for NCCO.
 The vector connecting the     
 two different unit-cells is shown, and the two orbitals     
 connected by the hopping as well.     
 }     
 \begin{tabular}{|c|c|c||c|c|c|}     
   \hline     
     orbitals & vector &  amplitude & orbitals & vector & amplitude  \\     
   \hline     
     $d_x$-$p_y$ &           (-2,0 ) &  -0.01 &      $p_y$-$d_x$ &           (-2,0 ) &  
 -0.01 \\
     $d_x$-$d_x$ &           (-1,-1) &  0.020 &      $d_x$-$p_x$ &           (-1,-1) &  
 -0.05 \\
     $d_x$-$p_y$ &           (-1,-1) &  0.050 &      $p_x$-$p_x$ &           (-1,-1) &  
 0.030 \\
     $p_y$-$p_y$ &           (-1,-1) &  0.030 &      $d_x$-$d_x$ &           (-1,0 ) &  
 -0.07 \\
     $d_x$-$p_x$ &           (-1,0 ) &  0.050 &      $d_x$-$p_y$ &           (-1,0 ) &  
 -1.16 \\
     $p_x$-$d_x$ &           (-1,0 ) &  0.050 &      $p_x$-$p_x$ &           (-1,0 ) &  
 -0.05 \\
     $p_x$-$p_y$ &           (-1,0 ) &  0.540 &      $p_y$-$d_x$ &           (-1,0 ) &  
 0.020 \\
     $p_y$-$p_y$ &           (-1,0 ) &  0.210 &      $d_x$-$d_x$ &           (-1,1 ) &  
 0.020 \\
     $d_x$-$p_y$ &           (-1,1 ) &  0.050 &      $p_x$-$d_x$ &           (-1,1 ) &  
 -0.05 \\
     $p_x$-$p_x$ &           (-1,1 ) &  0.030 &      $p_x$-$p_y$ &           (-1,1 ) &  
 -0.54 \\
     $p_y$-$p_x$ &           (-1,1 ) &  -0.01 &      $p_y$-$p_y$ &           (-1,1 ) &  
 0.030 \\
     $d_x$-$p_x$ &           (0 ,-2) &  0.010 &      $p_x$-$d_x$ &           (0 ,-2) &  
 0.010 \\
     $d_x$-$d_x$ &           (0 ,-1) &  -0.07 &      $d_x$-$p_x$ &           (0 ,-1) &  
 1.160 \\
     $d_x$-$p_y$ &           (0 ,-1) &  -0.05 &      $p_x$-$d_x$ &           (0 ,-1) &  
 -0.02 \\
     $p_x$-$p_x$ &           (0 ,-1) &  0.210 &      $p_y$-$d_x$ &           (0 ,-1) &  
 -0.05 \\
     $p_y$-$p_x$ &           (0 ,-1) &  0.540 &      $p_y$-$p_y$ &           (0 ,-1) &  
 -0.05 \\
     $d_x$-$d_x$ &           (0 ,0 ) &  2.450 &      $d_x$-$p_x$ &           (0 ,0 ) &  
 -1.16 \\
     $d_x$-$p_y$ &           (0 ,0 ) &  1.160 &      $p_x$-$d_x$ &           (0 ,0 ) &  
 -1.16 \\
     $p_x$-$p_x$ &           (0 ,0 ) &  0.840 &      $p_x$-$p_y$ &           (0 ,0 ) &  
 -0.54 \\
     $p_y$-$d_x$ &           (0 ,0 ) &  1.160 &      $p_y$-$p_x$ &           (0 ,0 ) &  
 -0.54 \\
     $p_y$-$p_y$ &           (0 ,0 ) &  0.840 &      $d_x$-$d_x$ &           (0 ,1 ) &  
 -0.07 \\
     $d_x$-$p_x$ &           (0 ,1 ) &  -0.02 &      $d_x$-$p_y$ &           (0 ,1 ) &  
 -0.05 \\
     $p_x$-$d_x$ &           (0 ,1 ) &  1.160 &      $p_x$-$p_x$ &           (0 ,1 ) &  
 0.210 \\
     $p_x$-$p_y$ &           (0 ,1 ) &  0.540 &      $p_y$-$d_x$ &           (0 ,1 ) &  
 -0.05 \\
     $p_y$-$p_y$ &           (0 ,1 ) &  -0.05 &      $d_x$-$p_x$ &           (0 ,2 ) &  
 0.010 \\
     $p_x$-$d_x$ &           (0 ,2 ) &  0.010 &      $d_x$-$d_x$ &           (1 ,-1) &  
 0.020 \\
     $d_x$-$p_x$ &           (1 ,-1) &  -0.05 &      $p_x$-$p_x$ &           (1 ,-1) &  
 0.030 \\
     $p_x$-$p_y$ &           (1 ,-1) &  -0.01 &      $p_y$-$d_x$ &           (1 ,-1) &  
 0.050 \\
     $p_y$-$p_x$ &           (1 ,-1) &  -0.54 &      $p_y$-$p_y$ &           (1 ,-1) &  
 0.030 \\
     $d_x$-$d_x$ &           (1 ,0 ) &  -0.07 &      $d_x$-$p_x$ &           (1 ,0 ) &  
 0.050 \\
     $d_x$-$p_y$ &           (1 ,0 ) &  0.020 &      $p_x$-$d_x$ &           (1 ,0 ) &  
 0.050 \\
     $p_x$-$p_x$ &           (1 ,0 ) &  -0.05 &      $p_y$-$d_x$ &           (1 ,0 ) &  
 -1.16 \\
     $p_y$-$p_x$ &           (1 ,0 ) &  0.540 &      $p_y$-$p_y$ &           (1 ,0 ) &  
 0.210 \\
     $d_x$-$d_x$ &           (1 ,1 ) &  0.020 &      $p_x$-$d_x$ &           (1 ,1 ) &  
 -0.05 \\
     $p_x$-$p_x$ &           (1 ,1 ) &  0.030 &      $p_y$-$d_x$ &           (1 ,1 ) &  
 0.050 \\
     $p_y$-$p_y$ &           (1 ,1 ) &  0.030 &      $d_x$-$p_y$ &           (2 ,0 ) &  
 -0.01 \\
     $p_y$-$d_x$ &           (2 ,0 ) &  -0.01 &      $p_y$-$p_y$ &           (2 ,2 ) &  
 0.000 \\
   \hline     
   \hline     
 \end{tabular}     
 \label{table:lda_param_ncco}     
 \end{table}     

%%%%%%%%%%%%%%%%%%%%%%%%%%%%%%%%%%%%%%%%%%%%%%%%%%%%%%%%%%%%%%%%
%%%%%%%%%%%%%%%%%%%%%%%%%%%%%%%%%%%%%%%%%%%%%%%%%%%%%%%%%%%%%%%%
%%%%%%%%%%%%%%%%%%%%%%%%%%%%%%%%%%%%%%%%%%%%%%%%%%%%%%%%%%%%%%%%
%%%%%%%%%%%%%%%%%%%%%%%%%%%%%%%%%%%%%%%%%%%%%%%%%%%%%%%%%%%%%%%%
%%%%%%%%%%%%%%%%%%%%%%%%%%%%%%%%%%%%%%%%%%%%%%%%%%%%%%%%%%%%%%%%
%%%%%%%%%%%%%%%%%%%%%%%%%%%%%%%%%%%%%%%%%%%%%%%%%%%%%%%%%%%%%%%%
%%%%%%%%%%%%%%%%%%%%%%%%%%%%%%%%%%%%%%%%%%%%%%%%%%%%%%%%%%%%%%%%
%%%%%%%%%%%%%%%%%%%%%%%%%%%%%%%%%%%%%%%%%%%%%%%%%%%%%%%%%%%%%%%%
%%%%%%%%%%%%%%%%%%%%%%%%%%%%%%%%%%%%%%%%%%%%%%%%%%%%%%%%%%%%%%%%
%%%%%%%%%%%%%%%%%%%%%%%%%%%%%%%%%%%%%%%%%%%%%%%%%%%%%%%%%%%%%%%%
%%%%%%%%%%%%%%%%%%%%%%%%%%%%%%%%%%%%%%%%%%%%%%%%%%%%%%%%%%%%%%%%

\bibliographystyle{prsty}

\end{document}